\documentclass[a4paper,fleqn]{cas-sc}
\usepackage[authoryear]{natbib}   
\usepackage{amsmath,amssymb}
\usepackage{graphicx}
\usepackage{booktabs}
\usepackage{multirow}
\usepackage{siunitx}
\usepackage{lipsum}
\usepackage{float}
\usepackage{placeins}
\usepackage{subcaption}
\usepackage{flafter}

\shorttitle{OpenMOC--OpenFOAM coupling with preCICE}
\shortauthors{S. S. W. Choo et al.}

\begin{document}
\let\WriteBookmarks\relax

\title[mode = title]{Development and Validation of a preCICE-Based OpenMOC–OpenFOAM Coupling Framework for Nuclear Reactor Analysis} 


\author[1]{Seth S. W. Choo}
\author[1]{L. K. Ang}
\author[2]{Zhe Chuan Tan}
\cormark[1]
\ead{tan.zc@nus.edu.sg}

\affiliation[1]{organization={Department of Engineering Product Development, Singapore University of Technology and Design},
                city={Singapore},
                postcode={487372},
                country={Singapore}}

\affiliation[2]{organization={Singapore Nuclear Research and Safety Institute},
                city={16 Prince George's Park},
                postcode={118415},
                country={Singapore}}

\cortext[cor1]{Corresponding author}

\begin{abstract}
To address the long computation time required by stochastic solvers, which are applied in the coupled neutronics-thermal hydraulics (N-TH) calculations, a faster N-TH coupling framework is developed based on the multiphysics coupling library preCICE, which enables data exchange between OpenMOC and OpenFOAM. 
As an alternative to commonly used stochastic solvers, like OpenMC, OpenMOC is used as a neutronics solver providing much faster steady-state solution in a Picard iteration. 
To realize this coupling strategy, OpenMC is used for an initial one-time computation cost for multi-group cross sections data generation that is used throughout OpenMOC's calculations. 
A preCICE library adapter for OpenMOC is also developed, allowing data exchange between neutronic power distributions and thermal hydraulic temperature fields. 
Our proposed coupling framework is validated through simulations of a simplified $2\times2$ fuel assembly, and a quartered geometry of $17\times17$ fuel assembly. 
The comparison shows good agreement with established referenced solutions, highlighting our proposed coupling framework for the steady-state N-TH coupling at a fraction of the time required by stochastic solver codes. 
This advantages could be extended for future applications where fast approximated solutions are required to address the change of N-TH coupling due to small parametric changes like in digital twin of nuclear reactor. 
\end{abstract}


\maketitle

\section{Introduction}
\label{sec:intro}

The power distribution in a nuclear reactor core is governed by neutron transport, while the temperatures that determine the material properties are governed by heat conduction and coolant flow. The fuel temperature alters the resonance absorption through Doppler broadening, and the moderator density and temperature alter neutron moderation. 
This nonlinear interaction causes neutronics and thermal hydraulics is intrinsically a coupled multiphysics problem, which requires expensive computational modeling approach.
Consequently, many prior works have explored different coupling neutronic and thermal hydraulic solvers, such as OFELIA: coupling OpenMC with FEniCSx (\cite{LOI2024113480}), OpenMC with FLUENT (\cite{zhang2021mcfluent}), RMC with FLUENT (\cite{ZHAO2025109724}), and OpenMC with OpenFOAM (\cite{ZHANG2026106495}) providing the coupling frameworks to address the exchange of data between the solvers and the required long multiphysics simulation time.

Most of these works involve stochastic neutronic solvers and Monte Carlo simulation of individual neutron histories, which estimate reaction rates as tallies averaged over many histories to solve the neutron transport equation (\cite{ROMANO201590}). 
Due to continuous-energy nuclear data and relatively simply geometry, Monte Carlo solvers are highly accurate. 
However, the large number of histories are required to maintain low statistical uncertainty of local quantities, such as pin-wise power at many axial levels.
This is computationally expensive, especially when the neutronics must be re-solved at every coupling iteration. This study aims to reduce the computational load on the neutronics side by implementing a deterministic solver, OpenMOC.

In terms of frameworks used, some studies opt for general-purpose frameworks that support multiphysics coupling, including MOOSE (\citep{FengKuaiyuan2025DoaM, StauffNicolasE.2024HMMo, MatthewsChristopher2021CMSo}) and MEDCoupling with the ICoCo interface (\citep{FengWenpei2021Doam}). MOOSE framework requires wrapping the solvers as MOOSE apps, whereas MEDCoupling, separates the coupled run supervisor (typically C3PO), and the data remapping library (MEDCoupling), whilst requiring solver codes to follow a fixed interface (ICoCo). 
Other neutronics-thermal hydraulics (N-TH) studies instead couple their solvers through custom scripts developed for the specific problem at hand. 
These implementations are well described in the literature showing tightly integrated implementations can be computationally efficient (\citep{zhang2021mcfluent}). However, such custom scripts are typically tailored to particular codes and geometries and are not distributed as general-purpose tools, making them less straightforward for other researchers to adopt and to improve for their own studies.
Thus, this study also aims to provide an open-source coupling workflow that is intuitive and can be easily extended to other geometries and codes, which is achieved by using the preCICE library (\cite{preCICEv2}) as a partitioned solver.

In this study, a coupling scheme between OpenMOC (deterministic) (\cite{BOYD201443}) and OpenFOAM (C++ Computational Fluid Dynamics libraries) (\cite{OpenFOAM}), taking advantage of the preCICE library as the partitioned solver, is built to provide an accessible alternative to N-TH multiphysics simulation. Advantages of this approach include the following:
\begin{enumerate}
    \item Coupling is defined through configuration files rather than modifications to the solvers, setting up a new coupled case requires minimal effort. 
    \item The framework achieves this at a modest computational cost, by introducing faster neutronics solving.
\end{enumerate}    
The results discussed in this study demonstrate the feasibility of extending our proposed framework to more complex geometries and to transient simulations, which will be presented in future papers.

\section{Methodology}
\label{sec:method}
This section introduces the underlying physics that is modeled for neutronics and thermal-hydraulics. 
It covers the details of the proposed coupling framework and the necessary extensions within the procedures listed below.

\subsection{Governing Equations}
\subsubsection{Neutronics}
Neutronic solvers aim to solve the steady-state Boltzmann neutron transport equation, Eq.~\eqref{eq:boltzmann}, written here in its eigenvalue $k$ form:

\begin{equation}
    \boldsymbol{\Omega}\cdot\nabla\psi(\mathbf{r},\boldsymbol{\Omega},E)
    + \Sigma_t(\mathbf{r},E)\,\psi(\mathbf{r},\boldsymbol{\Omega},E)
    = Q(\mathbf{r},\boldsymbol{\Omega},E)
    \label{eq:boltzmann}
\end{equation}
where $\psi$ is the angular neutron flux, $\Sigma_t$ is the total macroscopic cross section, and $Q$ is the combined scattering and fission source, with the fission term divided by the eigenvalue $k$.

Unlike OpenMC, which solves it stochastically, tracking every neutron history, OpenMOC instead solves Eq.~\eqref{eq:boltzmann} deterministically with the method of characteristics (MOC). The geometry is split into flat source regions (FSRs) of constant source and material properties, and the equation is integrated analytically along the characteristic tracks for a chosen azimujtal and polar quadrature (\citep{BOYD201443}).

OpenMOC is a multi-group code and cannot use continuous-energy data; its multi-group nuclear data has to be imported from an HDF5 file (\cite{BOYD201443}). These multi-group cross sections (MGXS) must be generated beforehand by condensing continuous-energy data over a representative neutron spectrum. In this work, they are generated with OpenMC, as described in Section (\ref{sec:openmc_mgxs}). 

In this study, the 3D tracks are defined by an azimuthal track spacing in the radial plane and a polar $z$-spacing, with a Gauss-Legendre polar quadrature (\cite{Gunow2017MC}). Transport sweeps are further accelerated with the implementation of Coarse Mesh Finite Difference (CMFD) (\cite{Gunow2019BEAVRS}).

\subsubsection{Thermal-hydraulics}
Early N-TH coupling efforts typically paired neutronics codes with one-dimensional system codes or two-dimensional subchannel codes, which represent the coolant as axial chanels with lumped cross-flow between them (\cite{WangBo2025Raao, SchmidtRodney2015Aafc, FengWenpei2021Doam}). As computational resources have grown, three-dimensional computational fluid dynamics (CFD) codes have been increasingly adopted (FLUENT (\cite{zhang2021mcfluent}), FLUENT (\cite{ZHAO2025109724}), OpenFOAM (\cite{FiorinaCarlo2015GanO}), OpenFOAM (\cite{ZHANG2026106495})), as they resolve local flow and temperature fields at the scale of the pin and subchannel. 

In this study, OpenFOAM is used as the thermal-hydraulics solver. Such CFD codes solve the conservation equations for mass, momentum, and energy over a discretized fluid domain. For the steady-state flow considered in this work, these are

\begin{align}
    \nabla\cdot(\rho\mathbf{u}) &= 0, \label{eq:mass}\\
    \nabla\cdot(\rho\mathbf{u}\mathbf{u}) &= -\nabla p
        + \nabla\cdot\boldsymbol{\tau} + \rho\mathbf{g}, \label{eq:momentum}\\
    \nabla\cdot(\rho c_p\mathbf{u}T) &= \nabla\cdot(k\nabla T) + q''', \label{eq:energy}
\end{align}
where Eq.~\eqref{eq:mass} is the continuity equation, for mass conservation; Eq.~\eqref{eq:momentum} is the momentum equation, in which momentum equates to pressure, viscous and gravitational forces; Eq.~\eqref{eq:energy} is the energy equation, in which heat carried by the flow equates to conduction and the volumetric heat source $q'''$. Here, $\rho$ is the density, $\mathbf{u}$ the velocity, $p$ the
pressure, $\boldsymbol{\tau}$ the viscous stress tensor, $\mathbf{g}$ the gravitational acceleration, $c_p$ the specific heat capacity, $k$ the thermal conductivity, $T$ the temperature, and $q'''$ the volumetric heat source. The OpenFOAM adapter provides modules for conjugate heat transfer (CHT), fluid-fluid and fluid-structure coupling, each of which supports compressible, incompressible and basic solver types. This study uses the CHT module: the coolant is solved with \texttt{bouyantSimpleFoam}, a steady-state variable-density solver, while the fuel and cladding are solved with \texttt{laplacianFoam}.

In the coolant region, only a single-phase fluid is considered and the energy Eq.~\eqref{eq:energy} is solved in enthalpy form:
\begin{equation}
\begin{split}
  \nabla\cdot(\rho\mathbf{u}h) + \nabla\cdot(\rho\mathbf{u}K)
     &= \nabla\cdot(\alpha_{\mathrm{eff}}\nabla h) \\
     &\quad + \rho\,\mathbf{u}\cdot\mathbf{g},
\end{split}
  \label{eq:energy_h}
\end{equation}
where $h$ is the specific enthalpy, $K=\tfrac12|\mathbf{u}|^{2}$ is the specific kinetic energy, and $\alpha_{\mathrm{eff}}$ is the effective thermal diffusivity. All equations here are solved with \texttt{buoyantSimpleFoam}.

In solid regions (fuel and cladding), there is no mass flow, so only Eq.~\eqref{eq:energy} is solved. In doing so, Eq.~\eqref{eq:energy} reduces to the steady heat conduction (without convection) of
\begin{equation}
    \nabla\cdot(k\nabla T) + q''' = 0. \label{eq:conduction}
\end{equation}
Here, it is solved with \texttt{laplacianFoam}, with $q'''$ the fission power density from the neutronics in the fuel and zero in the cladding, and a temperature-dependent conductivity $k(T)$ for UO$_2$.

\subsection{Neutronics-Thermal-hydraulics Feedback}
\label{sec:N-TH_feedback}
The governing equations of the two solvers (OpenMOC and OpenFOAM) are coupled through their material properties and source terms, with data being exchanged between both solvers. The neutronics solver depends on both the local fuel cross sections data and local moderator cross sections and density data, both of which are temperature dependent. The thermal-hydraulics solver depends on the volumetric power density, $q'''$, which becomes the heat source from fuel to clad region (solid-conduction). 

Neutronics depends on updated temperature and density fields. In the transport equation, Eq.~\eqref{eq:boltzmann}, the macroscopic cross sections depend on the local temperature, $T(\mathbf{r})$ and density, $\rho(\mathbf{r})$ of each material, creating two feedback mechanisms. The first feedback occurs as the fuel temperature rises, Doppler broadening widens the resonances in the cross sections of heavy nuclides such as $^{238}$U, reducing their self shielding. The effective multiplication factor can be written as:

\begin{equation}
  k_{\mathrm{eff}} = k_\infty\,P_{NL}
  = \frac{\nu\Sigma_f}{\Sigma_c+\Sigma_f}\,P_{NL},
  \label{eq:keff}
\end{equation}
where $\nu$ is the mean number of neutrons emitted per fission, $\Sigma_f$ and $\Sigma_c$ are the macroscopic fission and capture cross sections, and $P_{NL}$ ($0<P_{NL}\le1$) is the non-leakage probability. Broadening increases resonance capture in $^{238}$U ($\Sigma_c$) while leaving fission, which occurs mostly at thermal energies, nearly unchanged. $P_{NL}$ is also largely unaffected, so $k_{\mathrm{eff}}$ decreases as the fuel temperature increases.

The second feedback occurs as the coolant temperature rises, its density decreases, reducing neutron moderation. The relationship between density and neutron moderation can be described by
\begin{equation}
  \Sigma_x(\mathbf{r},E) = \sum_i N_i(\mathbf{r})\,\sigma_{x,i}(E),
  \label{eq:macro_xs}
\end{equation}

\begin{equation}
  N_{\mathrm{H_2O}}(\mathbf{r}) = \frac{\rho(\mathbf{r})\,N_A}{M_{\mathrm{H_2O}}},
  \label{eq:number_density}
\end{equation}
where $N_i$ is the number density of nuclide $i$, $\sigma_{x,i}(E)$ is its microscopic cross section for reaction $x$, $\rho$ is the water density, $N_A$ is Avogadro's number, and $M_{\mathrm{H_2O}}$ is the molar mass of water. Since $\rho$ is proportional to $N_i$, and hence to $\Sigma_x$ is, a decrease in coolant density ($\rho$) reduces the moderator cross sections ($\Sigma_{x,i}(E)$).

The thermal-hydraulics solver depends on the volumetric power density, $q'''$ which is computed from the fission reaction rate. Since the eigenvalue problem is homogeneous, $\phi$ is determined only up to a multiplicative constant, so the result is normalized to the total reactor power $P$:

\begin{equation}
  q'''(\mathbf{r})
  = \frac{P}{\displaystyle\int_V\!\int_0^\infty \kappa\,\Sigma_f\,\phi\,dE\,dV}
    \int_0^\infty \kappa\,\Sigma_f(\mathbf{r},E)\,\phi(\mathbf{r},E)\,dE,
  \label{eq:power_more}
\end{equation}
where $V$ is the fuel volume. The solution of Eq.~\eqref{eq:energy_h} gives the coolant temperature field and solution of Eq.~\eqref{eq:conduction} gives the fuel temperature field. For the calculations, coolant density is obtained from the coolant temperature using the IAPWS-IF97 formulation (\citep{iapws97}) at the system pressure.

Despite neutronics and thermal-hydraulics forming a tightly coupled, nonlinear system, it does not need to be solved as one complete system. Instead, the coupled solution is approximated with a partitioned approach: the two solvers run one after the other within a Gauss–Seidel Picard iteration. Details of this study's coupling approach and strategies to handle the feedback mechanisms are discussed in the following section.

\subsection{Coupling Framework}
In this study, the preCICE library is used as a partitioned coupling library to handle data exchange and mesh mapping between the neutronics and thermal–hydraulics solvers. Because each solver is connected to preCICE through an adapter, the coupling requires no changes to the solvers' source code, its internal physics and numerical methods remain unchanged, maintaining the physically accurate solutions. Since the solver codes communicate only through the exchanged fields, via their respective adapters, the proposed framework can still be implemented even if the solver codes are updated.

Previous works have extended the OpenFOAM-preCICE adapter (\citep{chourdakis2023openfoam}) with custom modules to exchange volumetric fields such as power density (\citep{ZHANG2026106495}).

\subsubsection{Data Exchanges and Mesh Mapping}
\label{sec:coupling_mesh}
\begin{figure*}[pos=t]
  \centering
  \includegraphics[width=\textwidth]{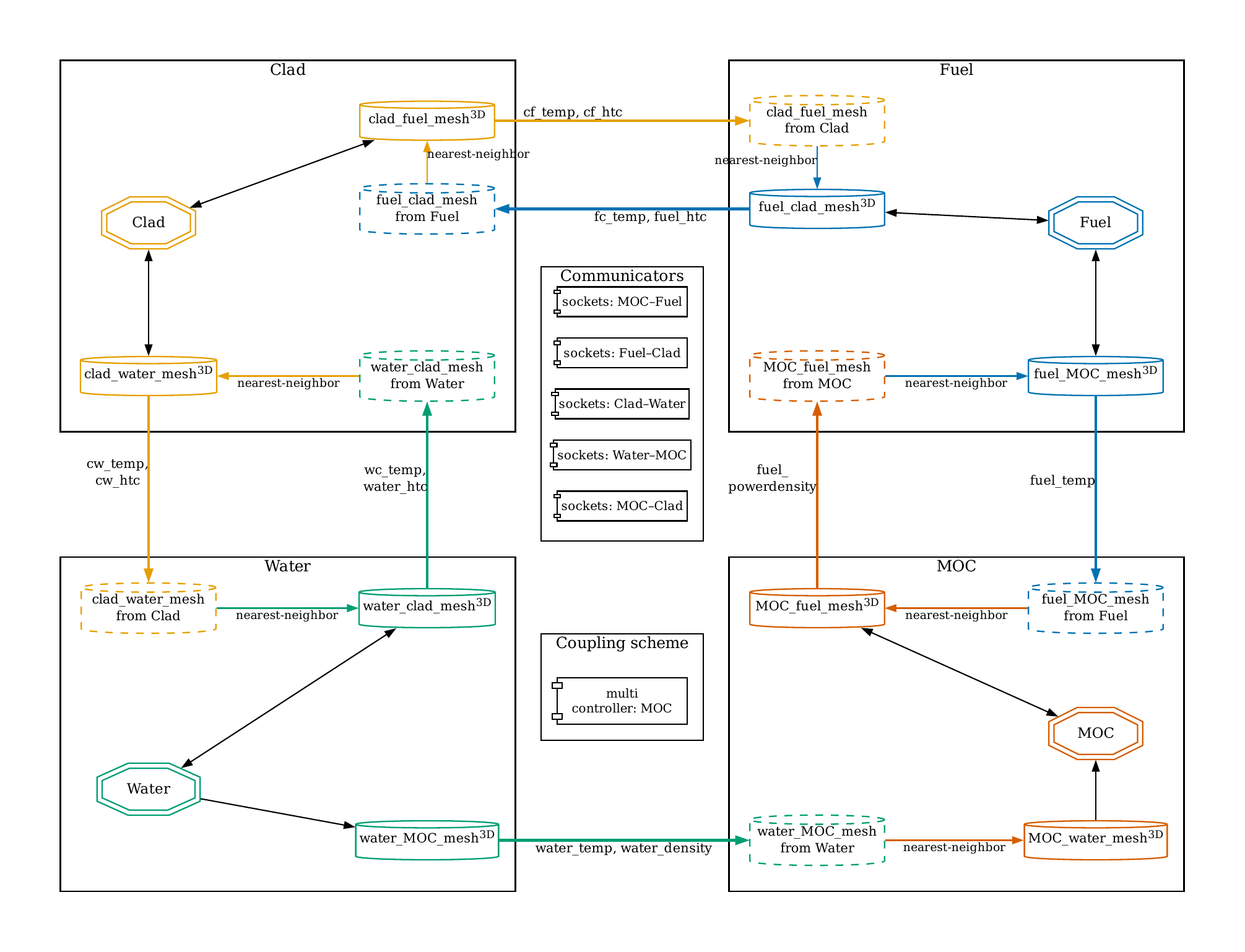}
  \caption{preCICE coupling configuration. Double octagons: participants;
    solid/dashed cylinders: provided/received meshes; thin coloured arrows:
    nearest-neighbour mappings; thick arrows: exchanged data, coloured by
    sender}
  \label{fig:coupling_topology}
\end{figure*}

The coupled system consists of four preCICE participants (solvers): three OpenFOAM solvers for the fuel, cladding and coolant regions (Fuel, Clad, Water) and the OpenMOC neutronics solver (MOC). The fuel and cladding are solved for heat conduction with \texttt{laplacianFoam}, and the coolant with the steady compressible solver \texttt{buoyantSimpleFoam}. The coupling configuration and exchanged data are shown in Fig.~\ref{fig:coupling_topology}.

Each participant provides its own coupling meshes and receives the meshes of the participants it exchanges data with. For example, with reference to Fig.~\ref{fig:coupling_topology}, the Clad solver provides \texttt{clad\_fuel\_mesh}, to send \texttt{cf\_temp} and \texttt{cf\_htc} for nearest neighbour mapping with particiapnt Fuel's corresponding \texttt{fuel\_clad\_mesh}. The internal thermal-hydraulics coupling between the solid and fluid regions is handled by the conjugate heat transfer (CHT) module of the OpenFOAM adapter. At the fuel-cladding and cladding-coolant interfaces, each side sends its interface temperature and heat transfer coefficient to the other, implementing a Robin-Robin coupling (\cite{SchollSebastian2018Sosc}). The neutronics coupling uses volumetric fields instead: MOC sends the \texttt{fuel\_powerdensity} to Fuel and receives the \texttt{fuel\_temp} from Fuel and the \texttt{water\_temp} and \texttt{water\_density} density from Water. All data are transferred with consistent nearest-neighbor mapping, defined on the reading participant. 

The four participants are coupled through a single multi-coupling scheme. Since MOC exchanges data with several participants, it is set as the controller. At the outer Picard iteration level, MOC starts the first solve, to produce the initial volumetric power density for the OpenFOAM solvers. Within each Picard iteration, OpenFOAM participants iterates over a series of OpenFOAM coupling windows, until the fuel-coolant interface (Fuel-Water or Clad-Water) converges. The number of windows required is governed by the coolant solver (Water), whose steady-state solution (\texttt{buoyantSimpleFoam}) requires many iterations, whereas the fuel conduction problem (\texttt{laplacianFoam}) converges much faster. The resulting temperature and density fields then update the cross sections for the next OpenMOC eigenvalue solve. These data exchanges, and subsequent solves are repeated until the convergence criteria is met. In this study, eigenvalue and the power distribution are set as the convergence criteria, given in Table. \ref{tab:picard_conv}. The change $\lvert k_{\mathrm{eff},n} - k_{\mathrm{eff},n-1} \rvert$ monitors global reactivity, the main neutronics result. However, from preliminary runs, $k_{\mathrm{eff}}$ converges before the power distribution. As such, max power change and RMS convergence criterion are introduced to ensure that all solvers converge. The tolerance values are chosen from preliminary runs.

\begin{table}[pos=H]
  \caption{Picard convergence criteria, evaluated between successive
    OpenMOC solves $n-1$ and $n$.}
  \label{tab:picard_conv}
  \setlength{\tabcolsep}{4pt}
  \begin{tabular}{@{}lll@{}}
    \toprule
    Quantity & Criterion & Tolerance \\
    \midrule
    Eigenvalue & $\lvert k_n - k_{n-1} \rvert$ & 5~pcm \\
    \addlinespace
    Power, RMS & $\mathrm{RMS}_b\big(\Delta q'''_b\big)\,/\,\overline{q'''}$ & $10^{-3}$ \\
    \bottomrule
  \end{tabular}
\end{table}

Under-relaxation is also applied to the power density to dampen oscillations between iterations and prevent the solution from diverging:

\begin{equation}
    q'''_{n+1} = (1-\alpha)\,q'''_{n} + \alpha\,\tilde{q}'''_{n+1},
    \label{eq:relaxation}
\end{equation}
where $\tilde{q}'''_{n+1}$ is the raw power density from the OpenMOC solve at iteration $n+1$, $q'''_{n}$ is the relaxed power density from the previous iteration, and $\alpha$ is the relaxation factor, set to 0.5 in this work.

Every iteration, each participant gives preCICE a list of vertex coordinates from their respective coupling meshes, and one value per vertex for each field. As such, there is a need for a uniform system that enables mesh mapping from FSRs in OpenMOC to CFD cells in OpenFOAM.

\begin{figure*}[pos=t]
  \centering
  \includegraphics[width=\textwidth]{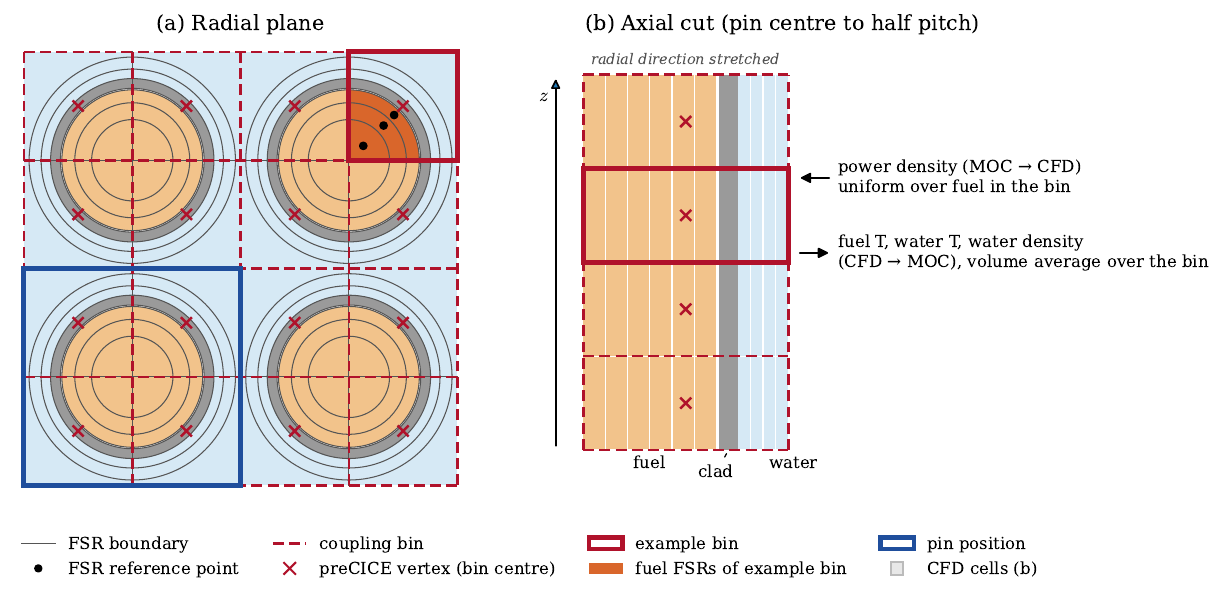}
  \caption{Coupling bins over the MOC geometry (schematic). (a)~Radial plane. (b)~Axial cut}
  \label{fig:coupling_mesh}
\end{figure*}

Both OpenMOC and OpenFOAM meshes use the same bins, shown as the red dashed squares in Fig.~\ref{fig:coupling_mesh}. Radially, each bin spans half a pin pitch, so every pin is covered by $2\times2$ bins; axially, each bin spans a set number of axial levels (depending on the model discussed in Section~\ref{sec:results_dis}). OpenMOC meshes have one vertex per bin, at the bin center, whereas OpenFOAM meshes use locations volumeCenters, so they have one vertex per CFD cell. For example, when \texttt{fuel\_powerdensity} data is transferred from MOC to CFD, each CFD cell center takes the value of the nearest bin center. Hence, all CFD cells in bin b gets a uniform power density of $q'''_b$. When \texttt{fuel\_temp}, \texttt{water\_temp} and \texttt{water\_density} data are transferred from CFD to MOC, each bin center takes the value of the nearest CFD cell center (only one cell out of all the cells in the bin). However, taking this raw CFD cell value would be a point sample, not a bin averaged value, so OpenFOAM first writes the bin average, $\bar{u}_b$ into every cell of the bin, allowing the map of the nearest cell value to always reflect that bin's average.

\begin{equation}
  \bar{u}_b = \frac{\sum_{c \in \mathcal{C}_b} u_c V_c}
                   {\sum_{c \in \mathcal{C}_b} V_c}
  \label{eq:bin_avg}
\end{equation}
where $u_c$ is the value of $T_f$, $T_w$ or $\rho_w$ in CFD cell $c$, $V_c$ is the volume of that cell, and $\mathcal{C}_b$ is the set of CFD cells whose centroids lie in bin $b$.

However, for this bin-based exchange to be unambiguous, no FSR or CFD cell may cross a bin edge. Both sides assign a region to a bin through a single point: OpenMOC uses the reference point of each FSR and the OpenFOAM participants use the centroid of each cell. A region that crosses a bin edge would be assigned entirely to one bin, although part of it lies in a neighbouring bin. Therefore, its power, temperature and density would be attributed to the wrong bin. Both meshes resolutions are designed with this limitation in mind. On the neutronics side, the four azimuthal sector cuts of each pin coincide with the radial bin edges, and the MOC axial levels are the bins themselves. On the thermal-hydraulics side, the O-grid blocks of each pin are split along the same center lines and the CFD mesh has exactly one cell layer per bin in the axial direction. Every FSR and every CFD cell therefore lies entirely within one bin, as shown in Fig.~\ref{fig:coupling_mesh}.

\subsubsection{OpenFOAM Adapter Extensions}
Instead of building a custom module, this study uses the Generic module introduced in version 1.4.0 of the adapter (\citep{chourdakis2023openfoam}) which couples arbitrary scalar and vector fields by name over volume or surface regions, without needing modifications to the adapter source code. This adapter's built-in modules each exchange a fixed set of quantities (Eg. CHT handles temperature, heat flux and heat-transfer coefficient), and none of the existing modules provide volumetric power density. The Generic module bypasses this restriction: in the adapter configuration, the preCICE data \texttt{fuel\_powerdensity} is linked to an OpenFOAM field by name, with one coupling vertex at the center of each fuel CFD cell.

Three extensions were made to the adapter's CHT module. First, the CHT module identifies each coupled quantity (e.g. sink temperature) from its preCICE data name, which ties the data name to the physical quantity. An optional \texttt{type} keyword was added to each data entry: when set, the module identifies the quantity from \texttt{type} instead, so that the data name can follow an arbitrary naming convention (e.g. \texttt{fc\_temp}, \texttt{cf\_htc} for the fuel-cladding interface). Second, for solid solvers without a thermophysical model (\texttt{solverType basic}, as used with \texttt{laplacianFoam}), the adapter computes the heat transfer coefficient from a single constant conductivity $k$. Since the conductivity of UO$_2$ varies strongly with temperature, a constant $k$ misrepresents the heat flux across the interface. An optional \texttt{kappaField} keyword was therefore added, which names a temperature-dependent conductivity field $k(T)$ updated by the solver, which is then used in the subsequent conduction solve. $k(T)$ is interpolated from the tabulated conductivity of fully dense UO$_2$ (\cite{fink1982uo2}) and corrected for pellet porosity. Third, in this study, the helium gap between fuel and the cladding is not meshed for both MOC and CFD. To address this, an optional \texttt{gapConductance} keyword adds a gap conductance $h_{\mathrm{gap}}$ in series on a chosen interface. Across the interface, heat conducts from the center of cell to the center of its adjacent cell. Within each half-cell, of thickness $d$ between the cell center and the interface face, Fourier's law gives

\begin{equation}
  q = \frac{k_1\,(T_{c,1} - T_{f})}{d_1} = \frac{k_2\,(T_{f} - T_{c,2})}{d_2},
  \label{eq:halfcell}
\end{equation}
where $q$ is the heat flux, $T_{c,1}$ and $T_{c,2}$ are the fuel and cladding cell-center temperatures and $T_f$ is the interface temperature. Each half-cell therefore acts as a thermal resistance per unit area and the two
act in series:

\begin{equation}
  q = \frac{T_{c,1} - T_{c,2}}{R_1 + R_2}
    = \frac{T_{c,1} - T_{c,2}}{d_1/k_1 + d_2/k_2}.
  \label{eq:robin_flux}
\end{equation}
The helium gap is not meshed, so it is represented by a third resistance in series,
$R_{\mathrm{gap}} = 1/h_{\mathrm{gap}}$:
\begin{equation}
  q = \frac{T_{c,1} - T_{c,2}}{R_1 + R_{\mathrm{gap}} + R_2}
    = \frac{T_{c,1} - T_{c,2}}{d_1/k_1 + 1/h_{\mathrm{gap}} + d_2/k_2}.
  \label{eq:gap_flux}
\end{equation}

With the optional \texttt{gapConductance} keyword, the thermal resistance of the un-meshed helium gap can be effectively modeled into the system.

\subsubsection{OpenMOC Adapter Development}
\label{sec:openmoc_mesh}
\begin{figure*}[pos=t]
  \centering
  \includegraphics[height=0.7\textheight]{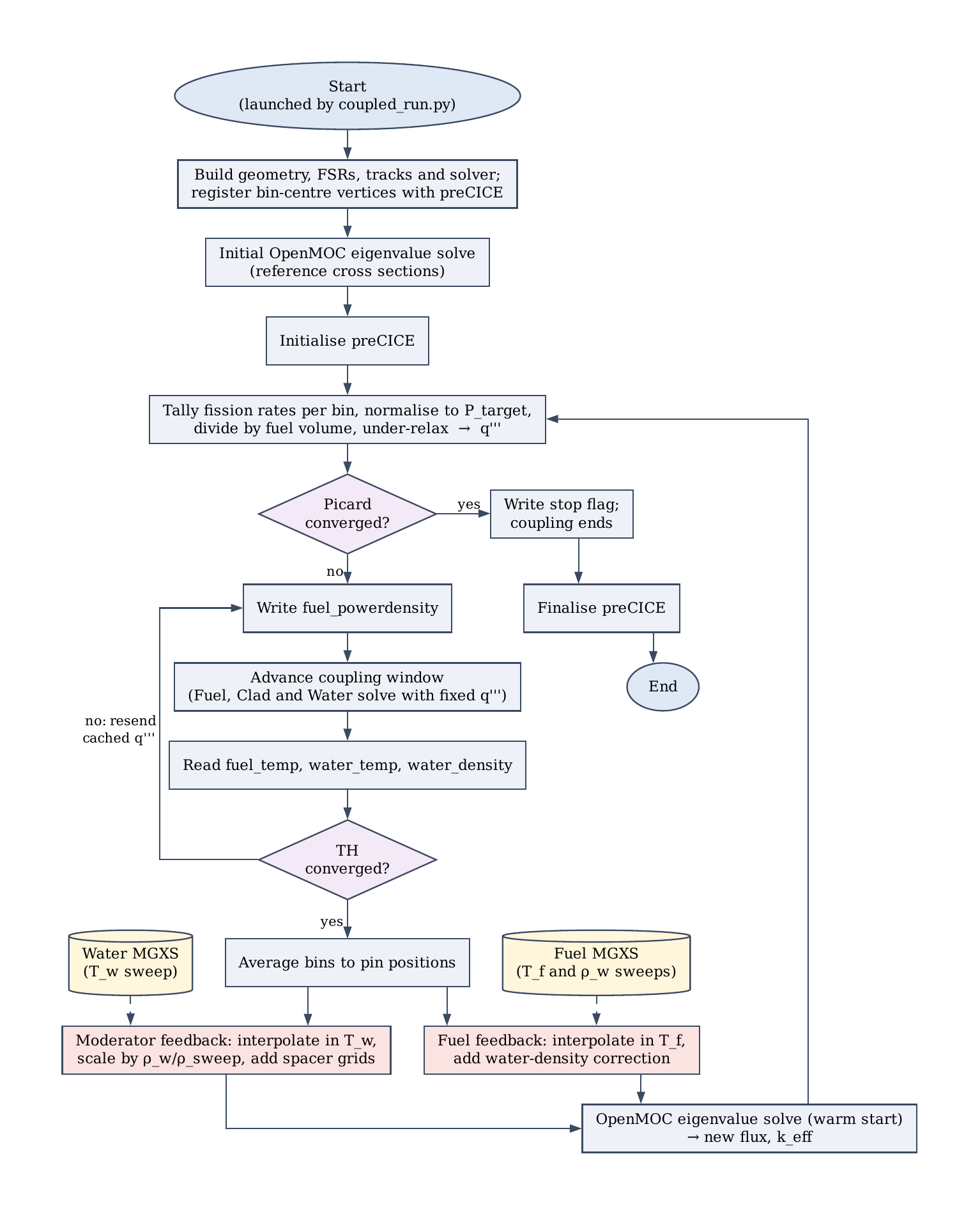}
  \caption{Workflow of the OpenMOC adapter}
  \label{fig:adapter_flowchart}
\end{figure*}

Similar to the existing OpenFOAM adapter, an adapter linking the preCICE library with OpenMOC was developed to handle the data exchange and mesh mapping, as no such adapter was previously available. The overall flow of the OpenMOC adapter is shown in Fig.~\ref{fig:adapter_flowchart}

Within each Picard iteration, the adapter converts the OpenMOC solution into the fuel power density sent to OpenFOAM (Fig.~\ref{fig:adapter_flowchart}). An eigenvalue solve gives the shape of the neutron flux but not its magnitude, so the fission rates are first tallied per bin and then scaled to the known power of the modeled domain. Each FSR $i$ is assigned to the bin that contains its reference point (Fig.~\ref{fig:coupling_mesh}), and the fission power of bin $b$ is

\begin{equation}
  P_b = E_f \sum_{i \in \mathcal{R}_b} \; \sum_{g=1}^{G}
        \Sigma_{f,g,i}\, \phi_{g,i}\, V_i,
  \label{eq:bin_power}
\end{equation}
where $\mathcal{R}_b$ is the set of FSRs whose reference points lie in bin $b$, $G$ is the number of energy groups, $\Sigma_{f,g,i}$ and $\phi_{g,i}$ are the fission cross section and scalar flux of FSR $i$ in energy group $g$, $V_i$ is the FSR volume, and $E_f = 3.2044\times10^{-11}$~J is the energy released per fission. The bin powers are then normalized so that their sum equals the target power, $P_{\mathrm{target}}$.

\begin{equation}
  \hat{P}_b = P_b \, \frac{P_{\mathrm{target}}}{\sum_{b'} P_{b'}}.
  \label{eq:normalise}
\end{equation}

Since a bin also contains gap, cladding and coolant, dividing $\hat{P}_b$ by the bin volume would dilute the heat source in the fuel. The power is therefore divided by the fuel volume of the bin only, given by the sum of the volumes of its fuel FSRs:

\begin{equation}
  V_{\mathrm{fuel},b} = \sum_{i \in \mathcal{F}_b} V_i,
  \qquad
  q'''_b = \frac{\hat{P}_b}{V_{\mathrm{fuel},b}},
  \label{eq:power_density}
\end{equation}
where $\mathcal{F}_b \subseteq \mathcal{R}_b$ is the set of fuel FSRs in bin $b$. Bins without fuel, such as those over the guide tubes, receive zero power density. 

\subsubsection{Thermophysical Properties Interpolation}
\label{sec:xs_interp}
\begin{figure*}[pos=t]
  \begin{subfigure}[b]{0.48\textwidth}
    \centering
    \includegraphics[width=0.888\linewidth]{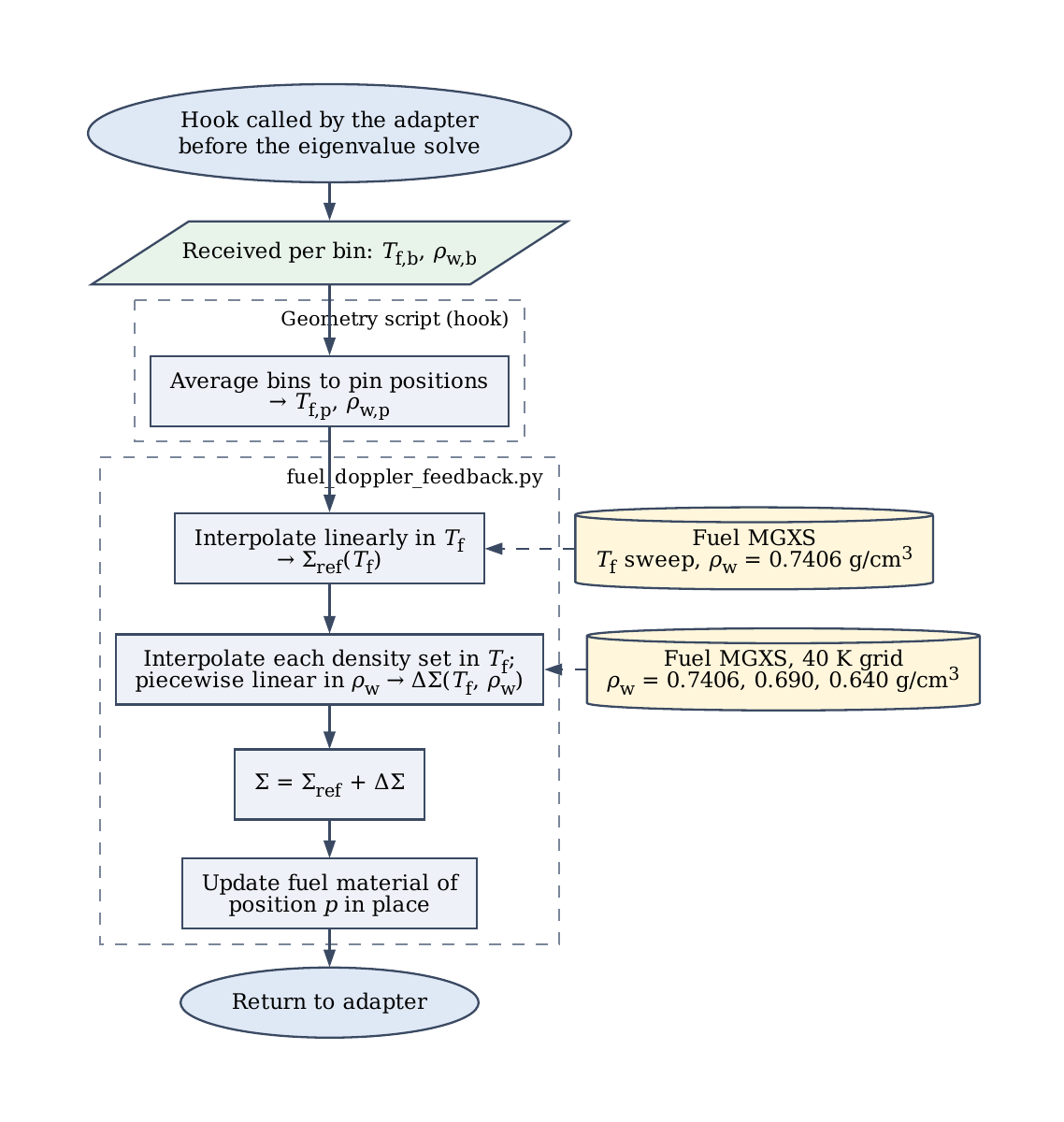}
    \caption{Fuel}
    \label{fig:fuel_feedback}
  \end{subfigure}\hfill
  \begin{subfigure}[b]{0.48\textwidth}
    \centering
    \includegraphics[width=\linewidth]{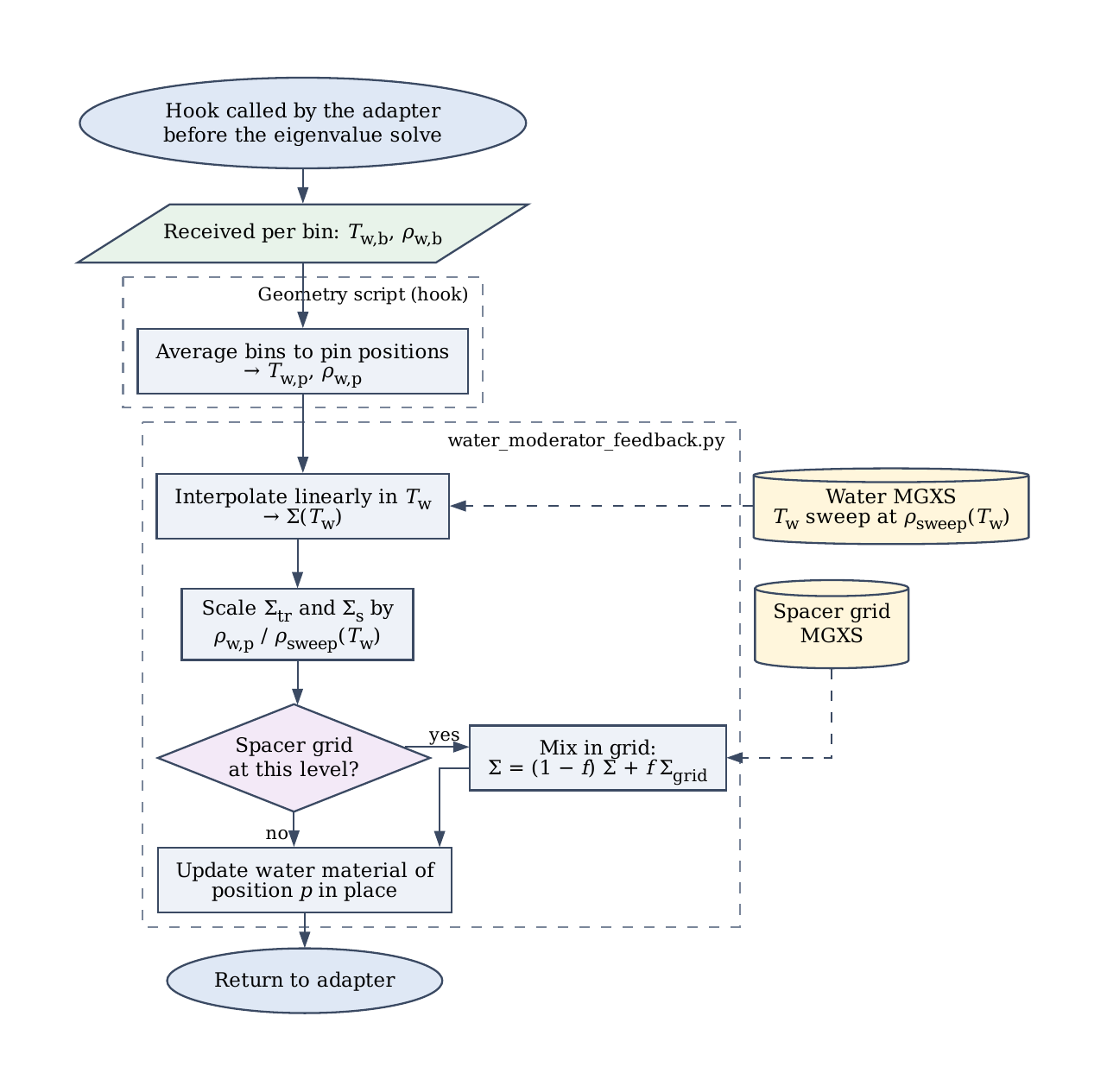}
    \caption{Coolant}
    \label{fig:water_feedback}
  \end{subfigure}
  \caption{Cross-section feedback: (a)~fuel, (b)~coolant}
  \label{fig:feedback}
\end{figure*}

OpenMOC solves the transport equation with fixed multigroup cross sections and has no internal temperature or density dependence. Thermal feedback is therefore implemented outside both OpenMOC and the adapter, in case-specific feedback modules called by the geometry script. Doing so gives the flexibility of changing the feedback modules without affecting the adapter. Before each eigenvalue solve, the adapter passes the received fuel temperature, coolant temperature and coolant density to these modules Fig.~\ref{fig:fuel_feedback} and Fig.~\ref{fig:water_feedback}, which update the cross sections of each pin position by interpolating in multi-group cross section (MGXS) tables precomputed over a range of thermal-hydraulic conditions. 

For the fuel feedback (Fig.~\ref{fig:fuel_feedback}), the cross sections, $\Sigma_{x,g}(T_f)$ at the pin-averaged fuel temperature $T_f$ are interpolated linearly between the two bracketing tabulated temperatures $T_k \le T_f < T_{k+1}$.

\begin{equation}
  \Sigma_{x,g}(T_f) = \Sigma_{x,g}(T_k)
    + \frac{T_f - T_k}{T_{k+1} - T_k}
      \big[\Sigma_{x,g}(T_{k+1}) - \Sigma_{x,g}(T_k)\big],
  \label{eq:doppler_interp}
\end{equation}
where $x$ denotes the reaction type and $g$ the energy group.

For the coolant (Fig.~\ref{fig:water_feedback}), the cross sections are interpolated linearly in the coolant temperature $T_w$ in the same way. Each tabulated coolant state was generated at a reference density $\rho_{\mathrm{sweep}}(T)$ from IAPWS-IF97 (\citep{iapws97}), which may differ from the density $\rho_w$ computed by the CFD solver. The transport and
scattering cross sections, which scale with the number density of the water molecules, are therefore corrected to first order by the density ratio:

\begin{equation}
  \Sigma_{x,g}(T_w,\rho_w) = \frac{\rho_w}{\rho_{\mathrm{sweep}}(T_w)}\;
                             \Sigma_{x,g}^{\mathrm{tab}}(T_w),
  \label{eq:density_ratio}
\end{equation}
where $\Sigma_{x,g}^{\mathrm{tab}}(T_w)$ is the cross section interpolated from the tables at the coolant temperature $T_w$ (Eq.~\eqref{eq:doppler_interp}), which
corresponds to the density $\rho_{\mathrm{sweep}}(T_w)$, and $\Sigma_{x,g}(T_w,\rho_w)$ is the cross section used in the transport solve, at the CFD temperature and density.

\subsubsection{OpenMC for MGXS Generation}
\label{sec:openmc_mgxs}
As mentioned above, OpenMOC requires the multi-group cross sections to be generated by an external solver. This study uses OpenMC, as this one-time initial cost before the coupled calculation. OpenMC's \texttt{openmc.mgxs} module has been shown to produce accurate multi-group cross sections for deterministic transport (\cite{Gunow2019BEAVRS}). The OpenMC model follows the same pin, guide tube and lattice dimensions as its OpenMOC geometry counterpart, applying the same reflective boundaries on all sides as well. The coolant is borated water at 1300 ppm. Continuous-energy data are taken from ENDF/B-VII.1, and each eigenvalue calculation uses 200 batches (50 inactive) of 20,000 particles.

Cross sections are tallied per material (fuel, helium gap, cladding and coolant) and condensed into the 16-group CASMO structure. A preliminary sweep of energy group structures was performed on a 2D $17\times17$ lattice at fuel temperatures of 600--1200~K, using continuous-energy OpenMC as the reference. The 16-group CASMO structure was selected as it offers a good balance between accuracy and cost for a PWR fuel assembly. For each material, the library contains the transport, fission, $\nu$-fission and $\nu$-scattering matrix cross sections and the fission spectrum $\chi$. Since OpenMOC assumes isotropic scattering, a P0 transport correction is applied, which folds the linearly anisotropic scattering component into the transport cross section and the diagonal of the scattering matrix.

In this study, the MGXS library is generated with all materials at the hot zero power temperature of 566.48~K, where the coolant density is 0.7406~g\,cm$^{-3}$ at the BEAVRS system pressure of 15.513~MPa (\cite{HorelikN.2013Bfea}). The tables used for the thermal feedback (Section~\ref{sec:xs_interp}) are generated by repeating this calculation with the state of one material changed at a time, shown in Table~\ref{tab:mgxs_sweeps}.

Each sweep varies the temperature of one material while the other is held at a fixed state (Table~\ref{tab:mgxs_sweeps}). All coolant states are evaluated at the BEAVRS system pressure of 15.513~MPa with the IAPWS-IF97 formulation~\citep{iapws97}, which relates the coolant temperature to its density.

For the fuel tables, the coolant is held at its reference state of 566.48~K and 0.7406~g\,cm$^{-3}$, and the fuel MGXS are generated over the fuel temperature range. Since the moderation of the coolant also affects the fuel cross sections, the fuel sweep is repeated with the coolant held at 0.690~g\,cm$^{-3}$ (589.3~K) and 0.640~g\,cm$^{-3}$ (606.7~K). These densities were chosen from the coolant density range observed in an initial coupled simulation, in which the fuel tables were generated only at the reference coolant state.

For the coolant tables, the fuel is held at 566.48~K, and IAPWS-IF97 is used for coolant densities below the saturation temperature of 618~K. Above 618~K, IAPWS-IF97 gives no liquid state data, so the linear $\rho(T)$ relation used by the CFD solver, which is itself a fit to IAPWS-IF97, is used instead. Two-phase water is not considered in this study since the coolant only boils in local regions of the reactor core, and the pressurized conditions of the reactor limits the effect of gaseous phase water. 

MGXS are generated over the coolant temperature range, each temperature paired with its density $\rho_{\mathrm{sweep}}(T_w)$. During the coupled calculation, the feedback modules interpolate between these tables at the temperatures and densities received from the CFD (Section~\ref{sec:xs_interp}).

\begin{table}[pos=H]
  \caption{MGXS tables generated with OpenMC.}
  \label{tab:mgxs_sweeps}
  \setlength{\tabcolsep}{4pt}
  \begin{tabular}{@{}llll@{}}
    \toprule
    Table & Varied & Range (step) & Held fixed \\
    \midrule
    Fuel    & $T_f$ & 565--1005~K (10~K)  & Coolant: 566.48~K, \\
            &       & 1025--1805~K (20~K) & \quad 0.7406~g\,cm$^{-3}$ \\
    \addlinespace
    Fuel,   & $T_f$ & 565--1805~K (40~K)  & Coolant: 589.3~K, \\
    density &       &                     & \quad 0.690~g\,cm$^{-3}$ \\
    corr.   &       &                     & Coolant: 606.7~K, \\
            &       &                     & \quad 0.640~g\,cm$^{-3}$ \\
    \addlinespace
    Coolant & $T_w$ & 565--705~K (10~K)   & Fuel: 566.48~K \\
            &       & at $\rho_{\mathrm{sweep}}(T_w)$ & \\
    \bottomrule
  \end{tabular}
\end{table}

\section{Results and Discussion}
\label{sec:results_dis}
This study's proposed N-TH coupling workflow is validated using a simplified $2\times2$ pressurized water reactor (PWR) fuel assembly, matching the model in (\cite{BullerwellLanceEdward2020AMCS}), and a $17\times17$ PWR quartered fuel assembly. The modeling details and results are discussed in the following sections. All coupled calculations were performed on a single desktop workstation with an AMD Ryzen~7 5700X processor (8 cores, 16 threads and 32~GB of memory). Solvers were run in parallel, and details of core allocation are shown in Table. \ref{tab:2x2_comparison} and Table. \ref{tab:coupling_17x17} for $2\times2$ and $17\times17$ models respectively.

\subsection{2 x 2 Simplified Lattice}

\subsubsection{Model Description}
\begin{figure*}[pos=tp]
  \begin{subfigure}[b]{0.48\textwidth}
    \centering
    \includegraphics[width=\linewidth,height=6cm,keepaspectratio]{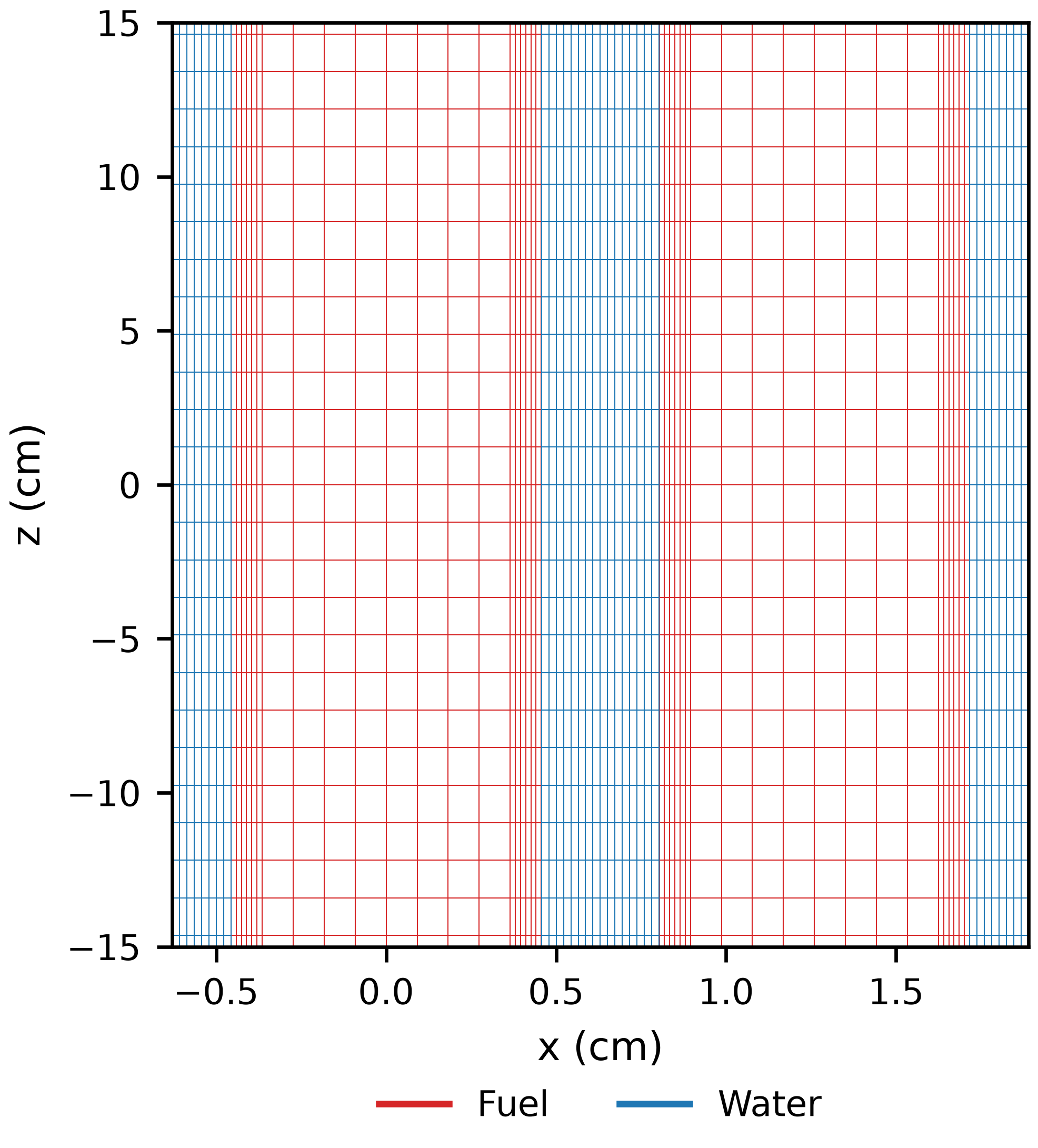}
    \caption{CFD, axial}
    \label{fig:2x2-cfd-mesh-axial}
  \end{subfigure}\hfill
  \begin{subfigure}[b]{0.48\textwidth}
    \centering
    \includegraphics[width=\linewidth,height=6cm,keepaspectratio]{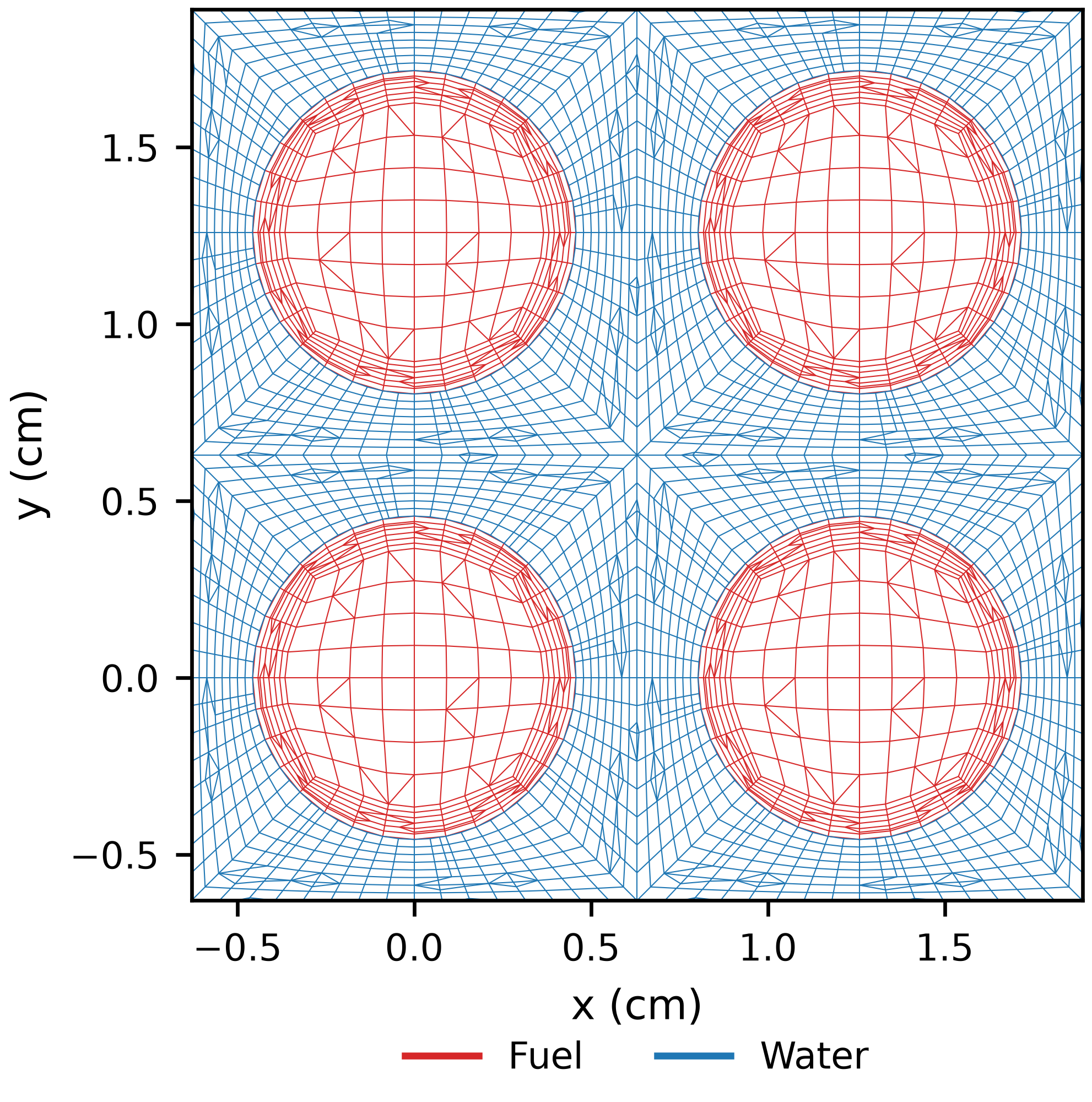}
    \caption{CFD, radial}
    \label{fig:2x2-cfd-mesh-radial}
  \end{subfigure}

  \vspace{0.5em}
  \begin{subfigure}[b]{0.48\textwidth}
    \centering
    \includegraphics[width=\linewidth,height=6cm,keepaspectratio]{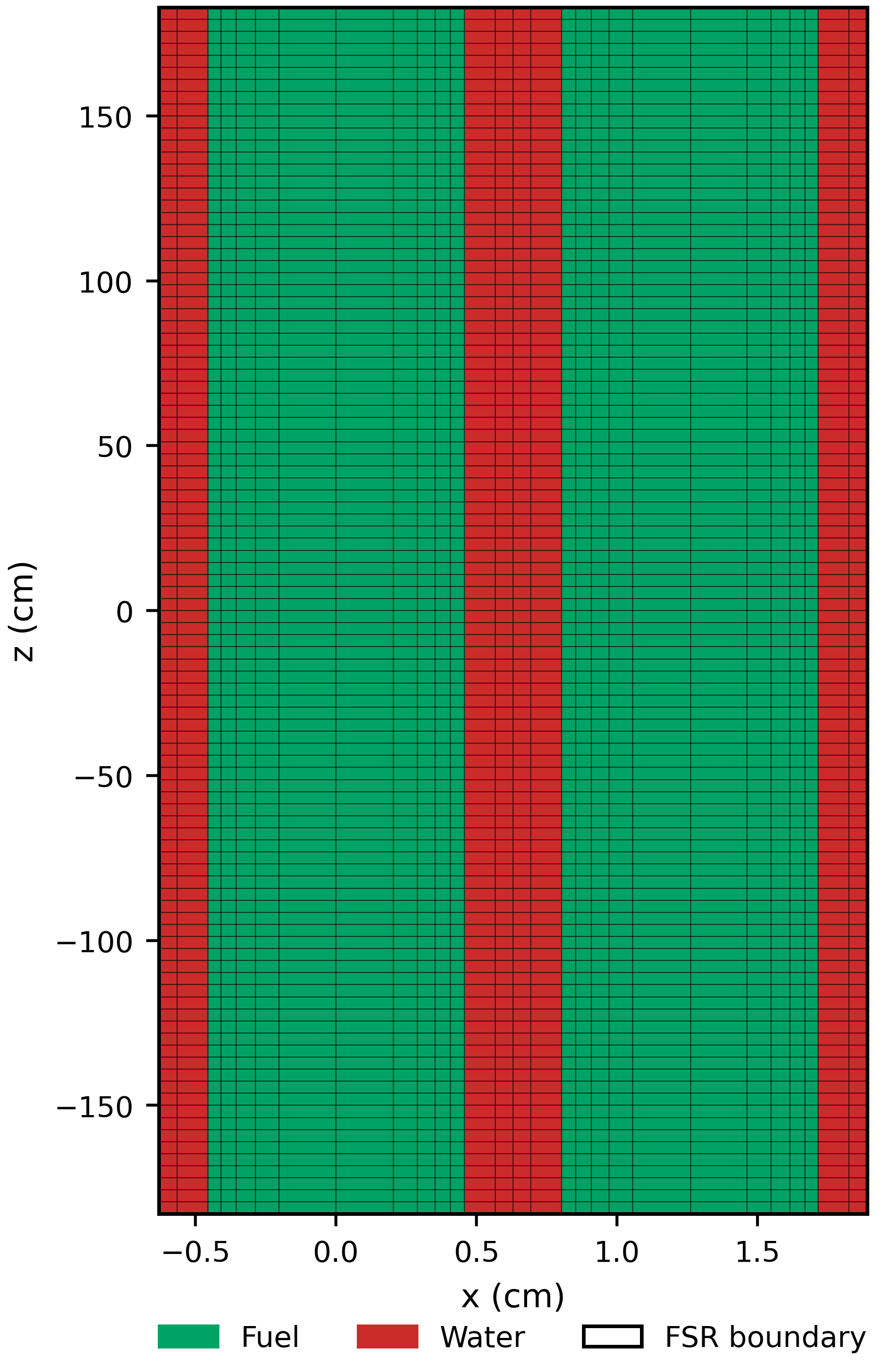}
    \caption{MOC FSRs, axial}
    \label{fig:2x2-moc-axial}
  \end{subfigure}\hfill
  \begin{subfigure}[b]{0.48\textwidth}
    \centering
    \includegraphics[width=\linewidth,height=6cm,keepaspectratio]{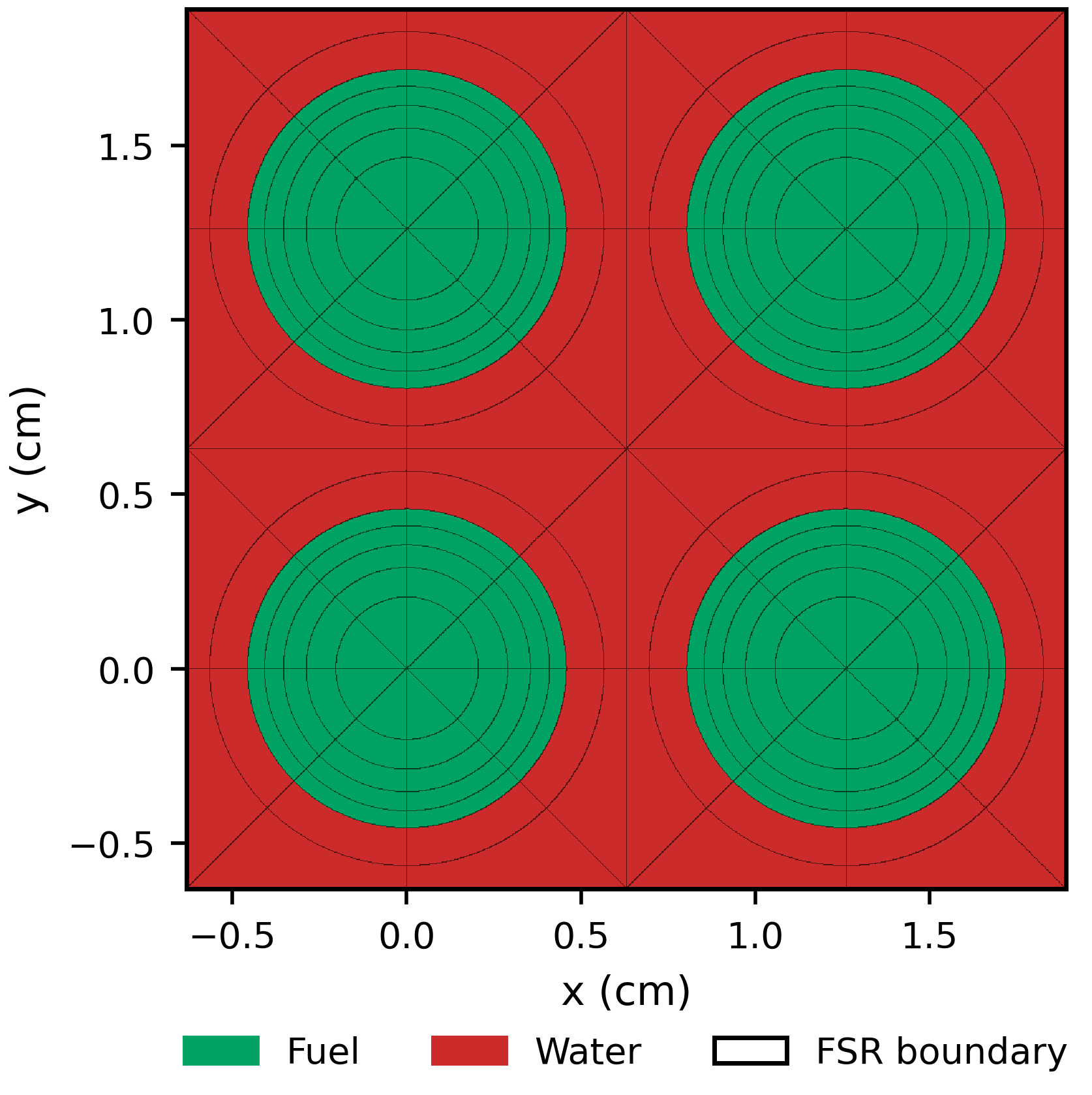}
    \caption{MOC FSRs, radial}
    \label{fig:2x2-moc-radial}
  \end{subfigure}
  \caption{Meshes of the 2$\times$2 pin case: (a)--(b)~CFD cells, (c)--(d)~OpenMOC flat source regions.}
  \label{fig:2x2-mesh}
\end{figure*}

To first verify the coupling workflow, a simplified $2\times2$ lattice of a PWR fuel assembly is modeled, and is compared to the benchmark by~\citep{BullerwellLanceEdward2020AMCS}. Only fuel and coolant regions are modeled, neglecting the cladding region. Geometric and material input parameters are listed in Table~\ref{tab:params_2x2} and Table~\ref{tab:props_2x2}. The geometric meshes are shown in Fig.~\ref{fig:2x2-mesh}. This case model uses $4\times4\times100$ coupling bins, with MOC having 100 axial levels (1 level per bin) and CFD having 300 axial cells (3 cells per bin).

\begin{table}[pos=htbp]
  \caption{Parameters of the simplified $2\times2$ fuel assembly model}
  \label{tab:params_2x2}
  \begin{tabular*}{\columnwidth}{@{\extracolsep{\fill}}ll@{}}
    \toprule
    Parameter & Value \\
    \midrule
    Fuel rod outer radius (cm)           & 0.4572 \\
    Lattice pitch (cm)                   & 1.25984 \\
    Axial height (cm)                    & 365.76 \\
    Fuel                                 & UO$_2$ \\
    Fuel enrichment                      & 3.1\% \\
    Coolant                              & Borated water \\
    Boron concentration (ppm)            & 1300 \\
    Inlet coolant temperature (K)        & 565 \\
    System pressure (MPa)                & 15.513 \\
    Mass flow rate (kg/s)                & 1.1097 \\
    Power (kW)                           & 267.781 \\
    \bottomrule
  \end{tabular*}

  \vspace{1.5em}

  \caption{Thermophysical properties of materials of $2\times2$ model}
  \label{tab:props_2x2}
  \begin{tabular*}{\columnwidth}{@{\extracolsep{\fill}}llll@{}}
    \toprule
    Material   & Density    & Conductivity           & Specific heat \\
               & (g/cm$^3$) & (W/m/K)                & (J/kg/K) \\
    \midrule
    UO$_2$     & 10.297     & (Fink) $k(T)\times0.911$ & 319 \\
    H$_2$O     & 0.7470     & 0.5806                 & 5286 \\
    \bottomrule
  \end{tabular*}
\end{table}

\clearpage
\subsubsection{Results}
\begin{figure}[pos=htbp]
  \centering
  \includegraphics[width=0.6\linewidth]{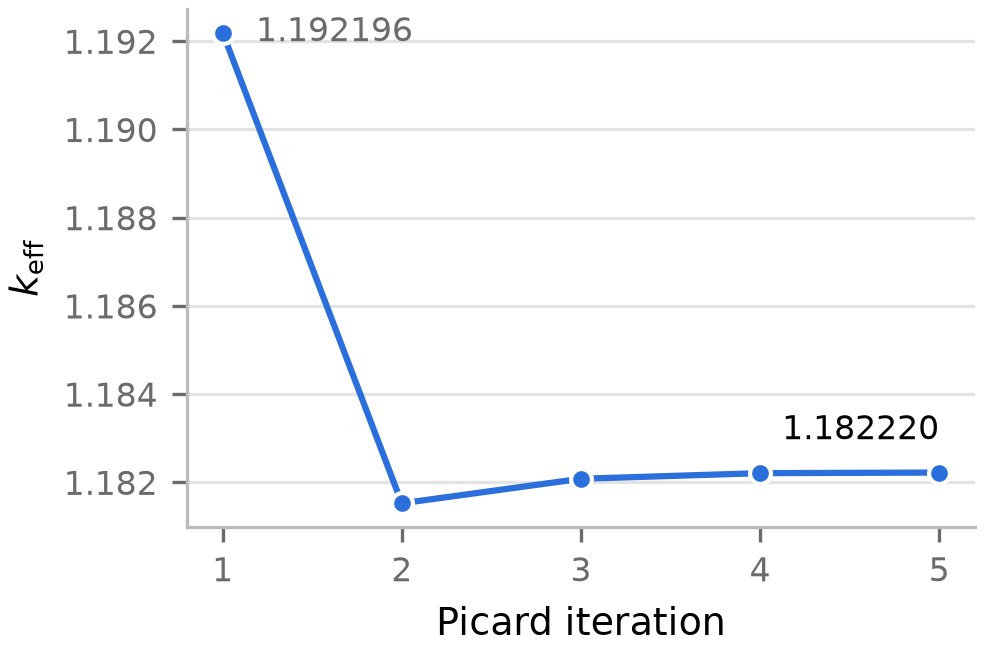}
  \caption{Variation of converged $k_\mathrm{eff}$ with iteration number for $2\times2$ case model}
  \label{fig:2x2_keff_v_iter}
\end{figure}

\begin{figure*}[pos=tp]
  \begin{subfigure}[b]{0.48\textwidth}
    \centering
    \includegraphics[width=\linewidth]{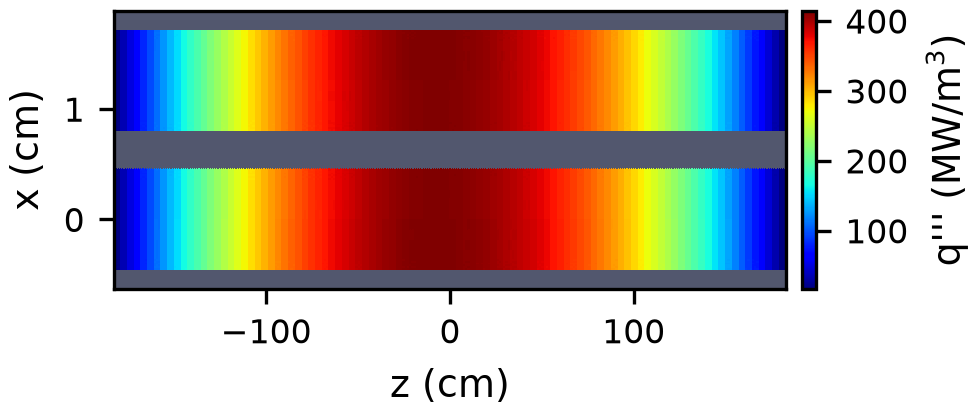}
    \caption{Calculated axial power at centerline}
    \label{fig:2x2_calc_power_axial_centerline}
  \end{subfigure}\hfill
  \begin{subfigure}[b]{0.48\textwidth}
    \centering
    \includegraphics[width=\linewidth]{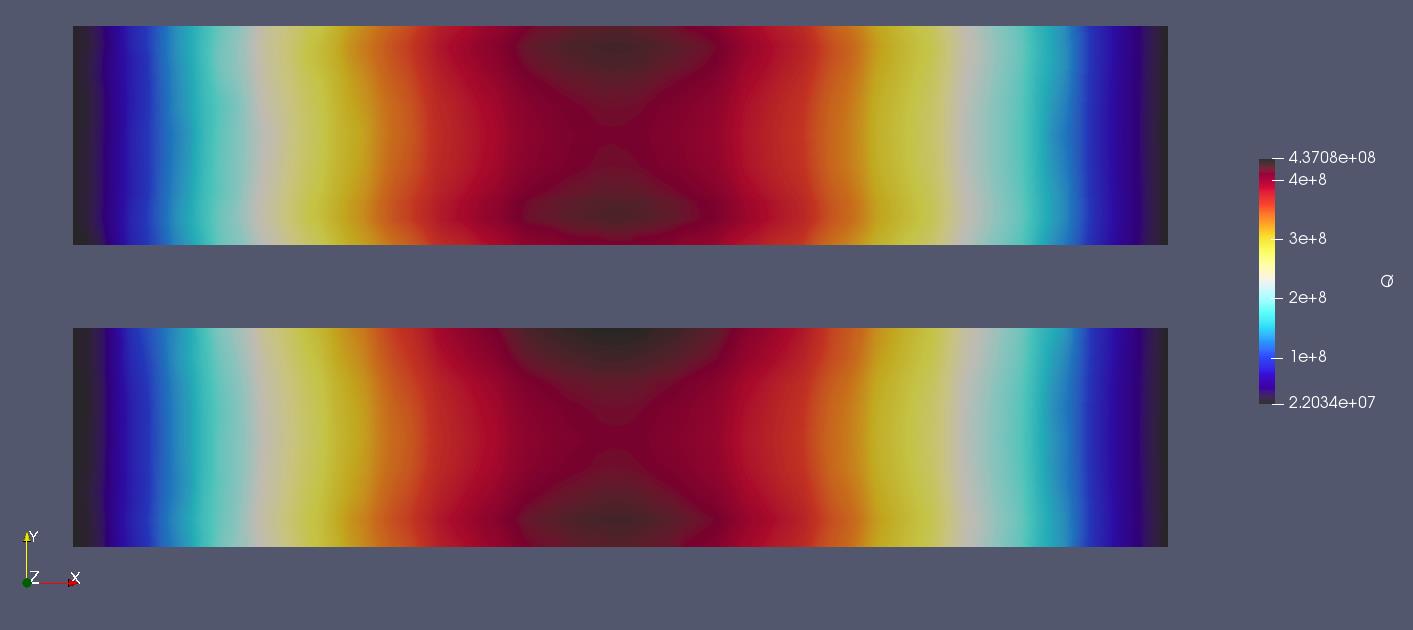}
    \caption{Reference axial power at centerline}
    \label{fig:2x2_ref_power_axial_centerline}
  \end{subfigure}
  \caption{Axial power at centerline of $2\times2$ model: (a)Calculated result, (b)Reference result}
  \label{fig:2x2_power_axial_centerline}
\end{figure*}

\begin{figure*}[pos=tp]
  \begin{subfigure}[b]{0.48\textwidth}
    \centering
    \includegraphics[width=\linewidth]{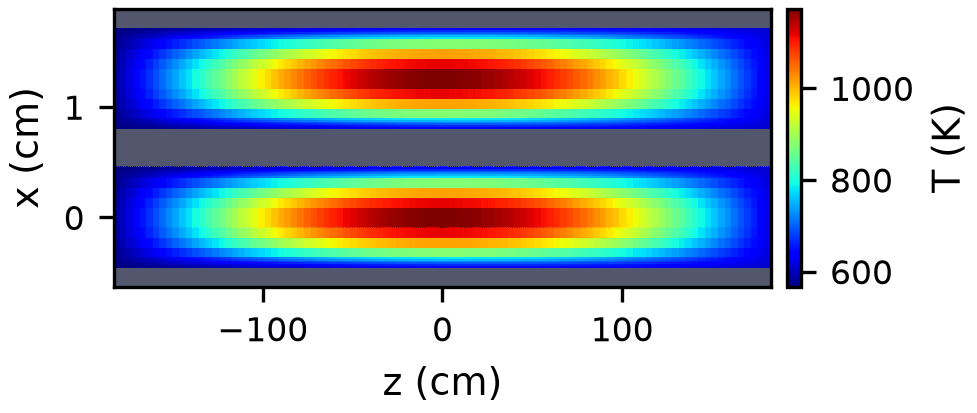}
    \caption{Calculated axial fuel temperature at centerline}
    \label{fig:2x2_calc_fuel_temp_axial}
  \end{subfigure}\hfill
  \begin{subfigure}[b]{0.48\textwidth}
    \centering
    \includegraphics[width=\linewidth]{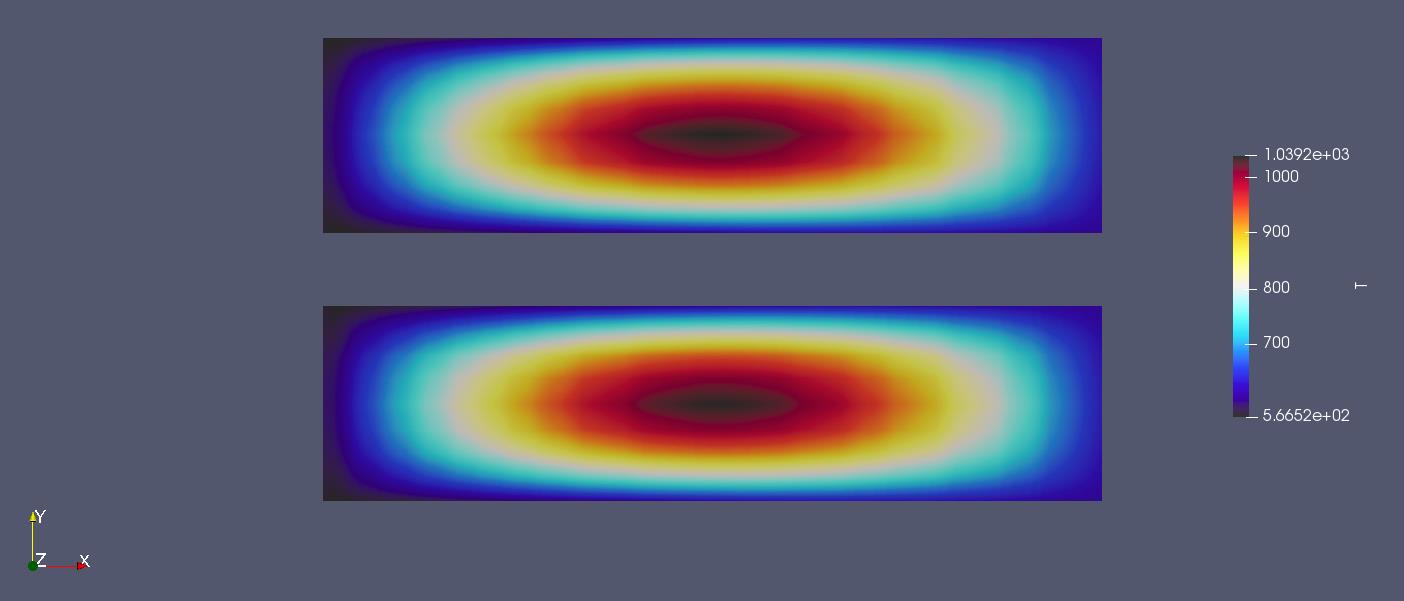}
    \caption{Reference axial fuel temperature at centerline}
    \label{fig:2x2_ref_fuel_temp_axial}
  \end{subfigure}
  \caption{Axial fuel temperature at centerline of $2\times2$ model: (a)Calculated result, (b)Reference result}
  \label{fig:2x2_fuel_temp_axial_centerline}
\end{figure*}

\begin{figure*}[pos=tp]
  \begin{subfigure}[b]{0.48\textwidth}
    \centering
    \includegraphics[width=\linewidth]{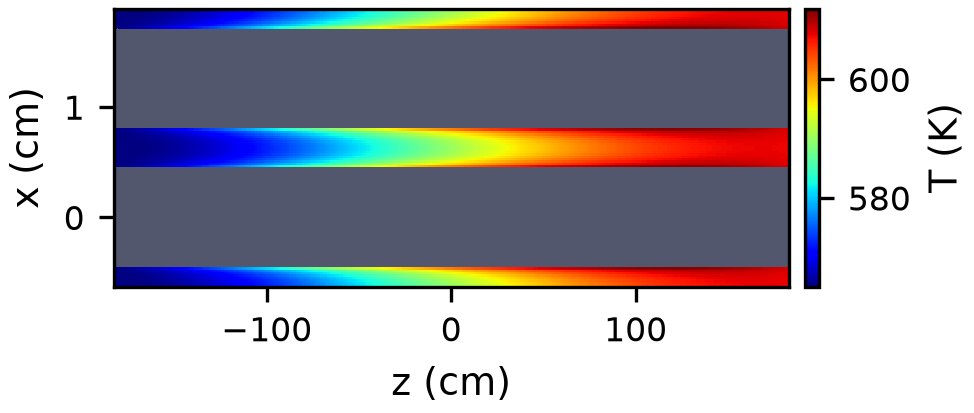}
    \caption{Calculated axial moderator temperature}
    \label{fig:2x2_calc_water_temp_axial}
  \end{subfigure}\hfill
  \begin{subfigure}[b]{0.48\textwidth}
    \centering
    \includegraphics[width=\linewidth]{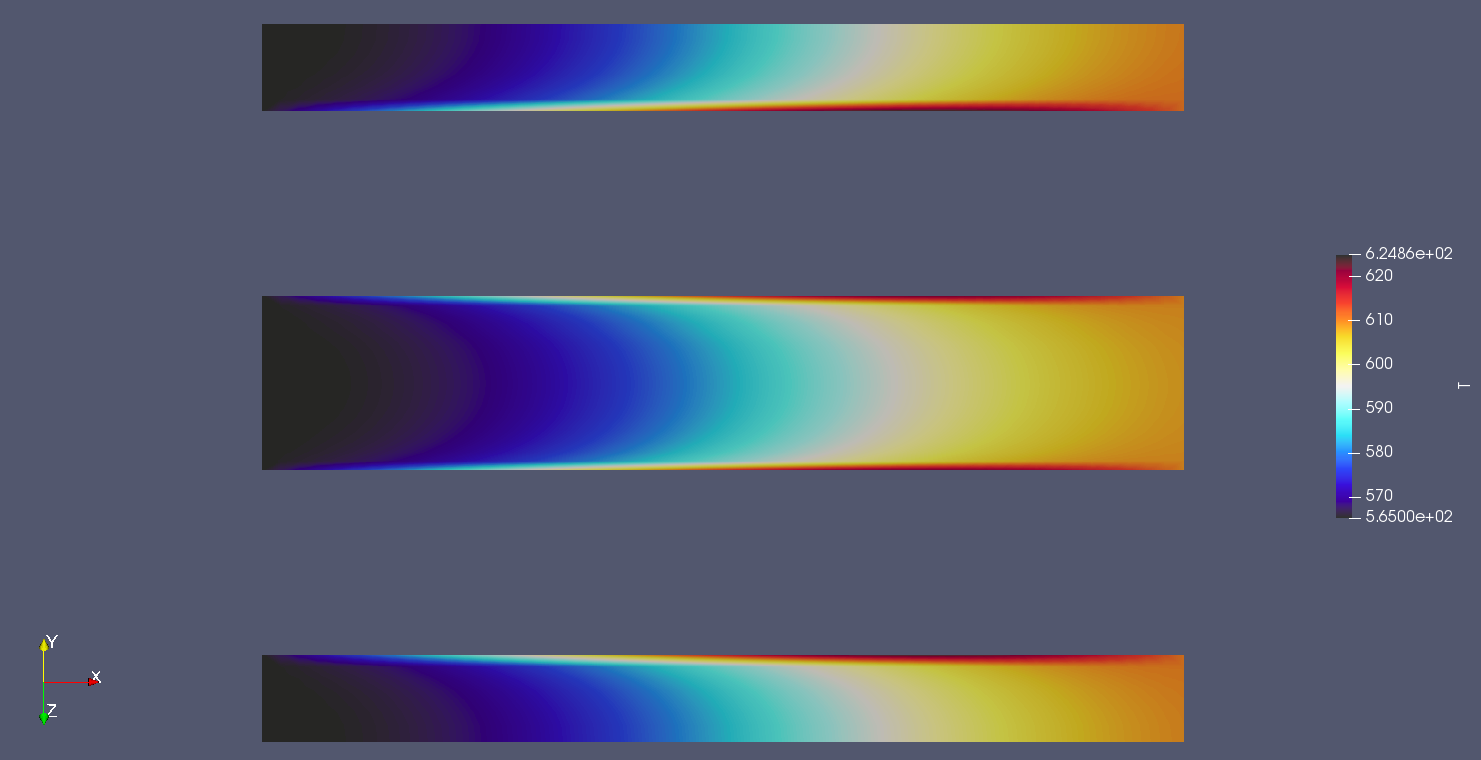}
    \caption{Reference axial moderator temperature}
    \label{fig:2x2_ref_water_temp_axial}
  \end{subfigure}
  \caption{Axial moderator temperature of $2\times2$ model: (a)Calculated result, (b)Reference result}
  \label{fig:2x2_water_temp_axial}
\end{figure*}

Fig.~\ref{fig:2x2_keff_v_iter} shows the change in $k_\mathrm{eff}$ over the iteration number. The solution approaches convergence within the third iteration, converging at a $k_\mathrm{eff}$ value of 1.18222, which is in agreement with the reference $k_\mathrm{eff}$ value of 1.18396 (\cite{BullerwellLanceEdward2020AMCS}). The offset can be attributed to the 16-group cross sections, compared to the continuous energy cross sections data that (\cite{BullerwellLanceEdward2020AMCS}) uses. Table.~\ref{tab:2x2_comparison} gives the full breakdown of values compared to (~\cite{BullerwellLanceEdward2020AMCS}). Fig.~\ref{fig:2x2_power_axial_centerline}, Fig.~\ref{fig:2x2_fuel_temp_axial_centerline} and Fig.~\ref{fig:2x2_water_temp_axial} shows the comparisons of power, fuel temperature and moderator temperature distributions.

\begin{table}[pos=htbp]
  \caption{Comparison of the $2\times2$ lattice results with
    \citet{BullerwellLanceEdward2020AMCS}. The present results hold the coolant
    density in the neutronics at 0.69572~g/cm$^3$.}
  \label{tab:2x2_comparison}
  \small
  \begin{tabular*}{\columnwidth}{@{\extracolsep{\fill}}lccc@{}}
    \toprule
    Quantity & Present & Reference & Diff. \\
    \midrule
    \multicolumn{4}{@{}l}{\emph{Neutronics}} \\
    $k_\mathrm{eff}$                 & 1.182220 & 1.18396$^{a}$ & $-174$~pcm \\
    Avg.\ $q'''$ (MW/m$^3$)          & 278.7    & 278.96        & $-0.09$\% \\
    Peak $q'''$ (MW/m$^3$)           & 411.5    & 437.08        & $-5.85$\% \\
    \midrule
    \multicolumn{4}{@{}l}{\emph{Thermal-hydraulics}} \\
    Coolant outlet $T$ (K)           & 605.9    & 609.79        & $-3.9$~K \\
    Peak coolant $T$ (K)             & 611.9    & 624.86        & $-13.0$~K \\
    \midrule
    \multicolumn{4}{@{}l}{\emph{Fuel temperature}} \\
    Avg.\ fuel $T$ (K)               & 777.7    & 742.66        & $+35.0$~K \\
    \bottomrule
    \multicolumn{4}{@{}l}{\footnotesize $^{a}$\,$\pm4$~pcm.}
  \end{tabular*}
\end{table}

Table.~\ref{tab:2x2_runtime} shows the core allocation for this $2\times2$ coupled calculation, and total run times are compared to Table 2 in (\cite{BullerwellLanceEdward2020AMCS}). The total run time of this coupled simulation (including one-time computation cost of MGXS generation) at 8 cores, is 02:42:44, which is considerably faster than the 17:33:40 from Picard iteration with adaptive batching used by (\cite{BullerwellLanceEdward2020AMCS}).

\begin{table}[pos=htbp]
  \caption{Computational time of the $2\times2$ coupled calculation, by participant,
    compared with \citet{BullerwellLanceEdward2020AMCS}.}
  \label{tab:2x2_runtime}
  \small
  \begin{tabular*}{\columnwidth}{@{\extracolsep{\fill}}lcc@{}}
    \toprule
    Case & Cores & Time (hh:mm:ss) \\
    \midrule
    \multicolumn{3}{@{}l}{\emph{Present work (OpenMOC/OpenFOAM)$^{c}$}} \\
    \quad MGXS generation (OpenMC)$^{d}$    & 8$^{a}$ & 01:51:29 \\
    \quad Fuel (\texttt{laplacianFoam})      & 1       & 00:19:09 \\
    \quad Water (\texttt{buoyantSimpleFoam}) & 5       & 00:43:52 \\
    \quad MOC (OpenMOC)                      & 2$^{a}$ & 00:07:22 \\
    \quad Coupled run (wall clock)           & 8       & 00:51:15 \\
    \quad Total (wall clock)$^{e}$           & 8       & 02:42:44 \\
    \midrule
    \multicolumn{3}{@{}l}{\emph{\citet{BullerwellLanceEdward2020AMCS} (OpenMC/OpenFOAM)$^{b}$}} \\
    \quad Picard                             & --      & $>$240:00:00 \\
    \quad Picard, power relaxation           & --      & 19:05:29 \\
    \quad Picard, adaptive batching          & --      & 17:33:40 \\
    \bottomrule
    \multicolumn{3}{@{}p{\columnwidth}@{}}{\footnotesize
      $^{a}$\,OpenMP threads.
      $^{b}$\,Wall-clock time to reach a power convergence of $10^{-3}$ (Table~2 of the source).
      $^{c}$\,Coupled calculation with coolant density feedback, started from a cold state.
      $^{d}$\,Reference library, fuel and water temperature sweeps, and fuel tables at two coolant densities,
      run once before the coupled calculation.
      $^{e}$\,MGXS generation plus run before coupled simulation.}
  \end{tabular*}
\end{table}
\clearpage

\subsection{17 x 17 PWR Fuel Assembly}

\subsubsection{Model Description}
\begin{figure}[pos=htbp]
  \begin{subfigure}[b]{0.48\textwidth}
    \centering
    \includegraphics[width=\linewidth,height=6cm,keepaspectratio]{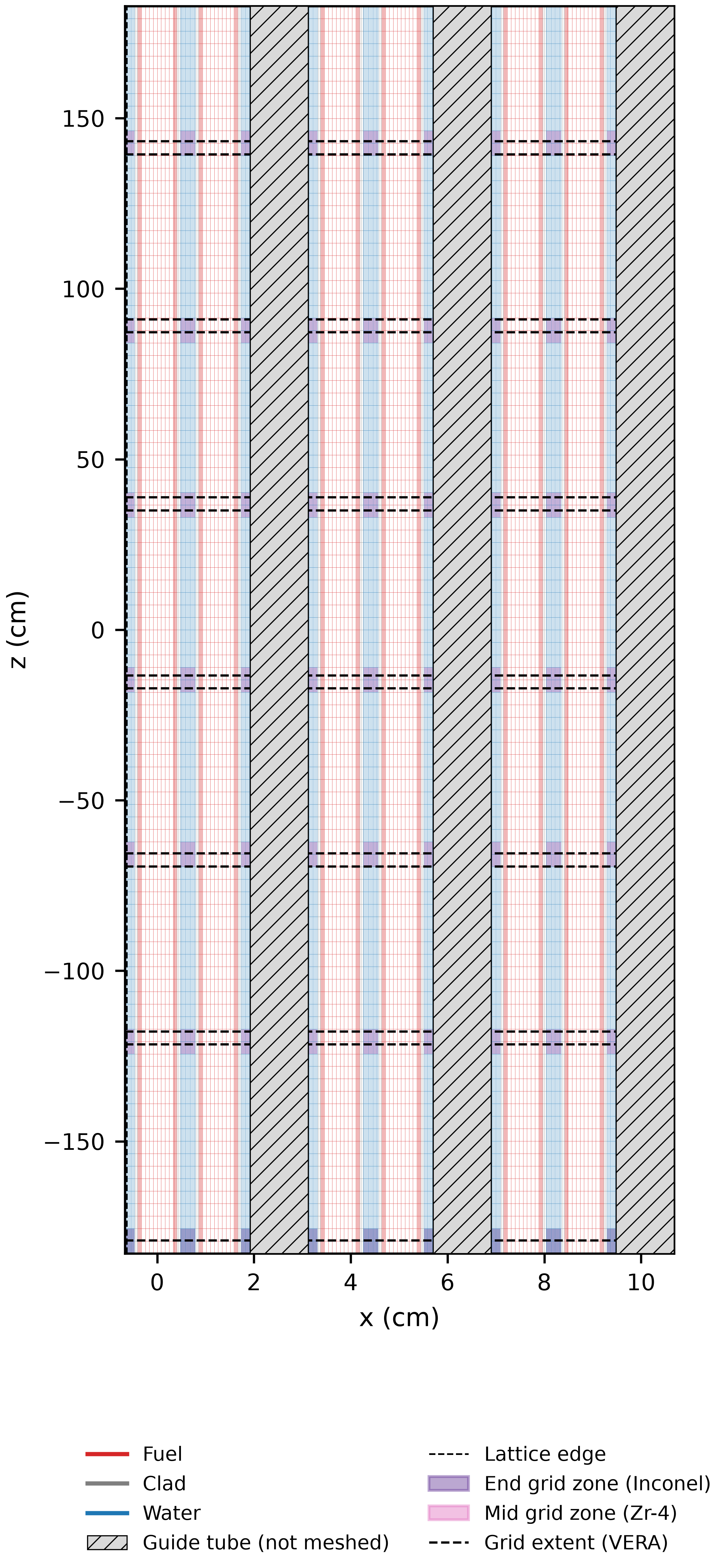}
    \caption{CFD, axial}
    \label{fig:17x17-cfd-mesh-axial}
  \end{subfigure}\hfill
  \begin{subfigure}[b]{0.48\textwidth}
    \centering
    \includegraphics[width=\linewidth,height=6cm,keepaspectratio]{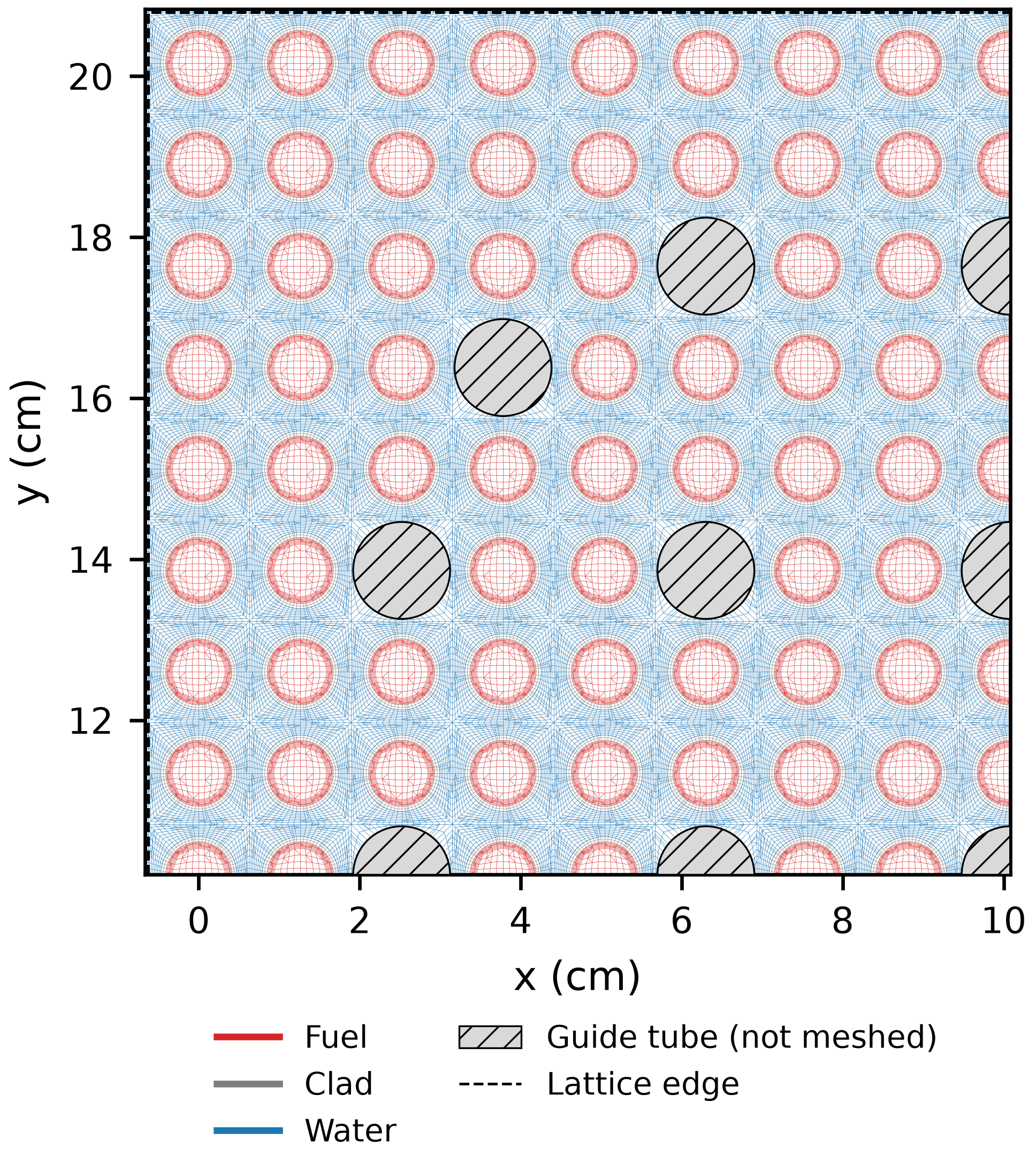}
    \caption{CFD, radial}
    \label{fig:17x17-cfd-mesh-radial}
  \end{subfigure}

  \vspace{0.5em}
  \begin{subfigure}[b]{0.48\textwidth}
    \centering
    \includegraphics[width=\linewidth,height=6cm,keepaspectratio]{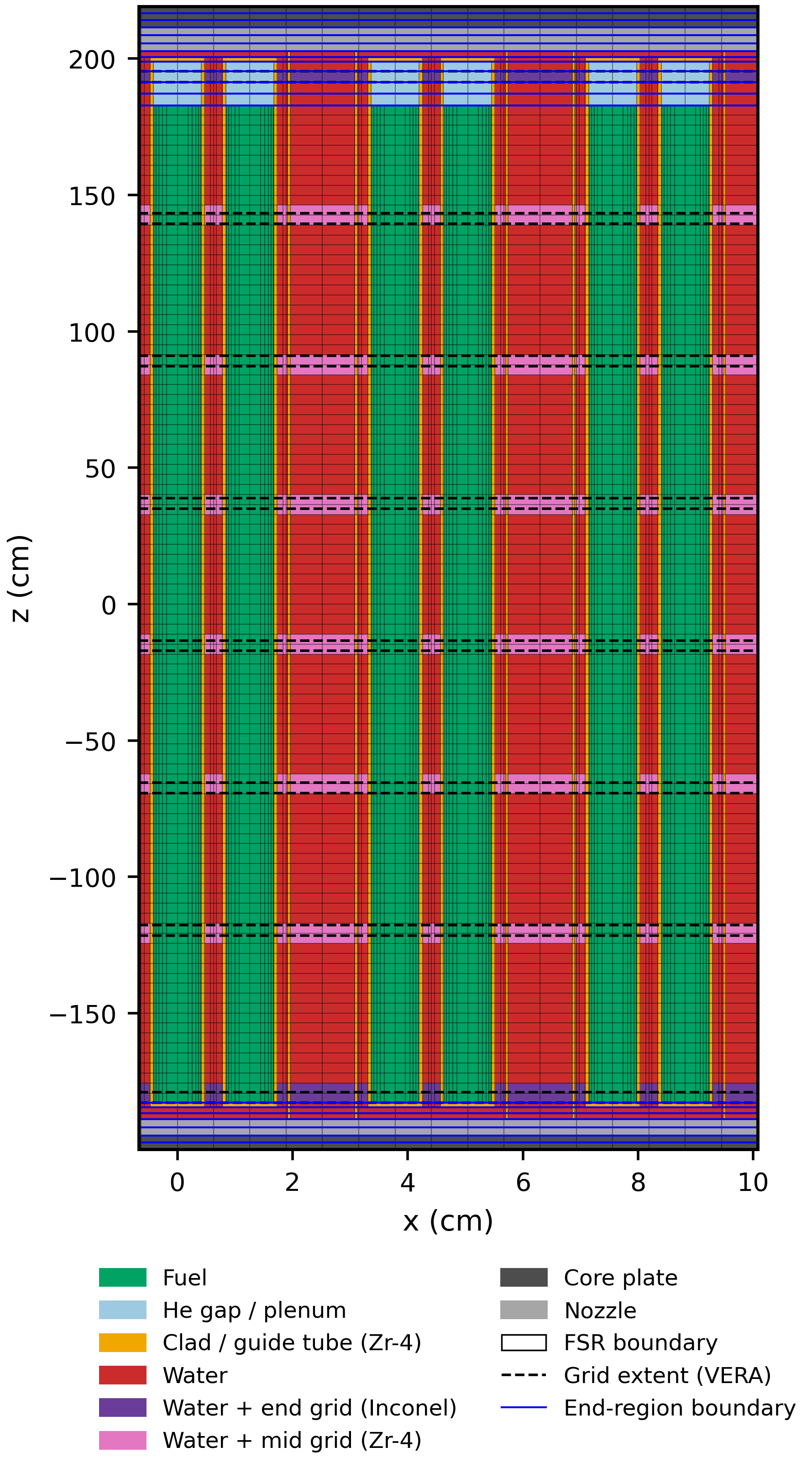}
    \caption{MOC FSRs, axial}
    \label{fig:17x17-moc-axial}
  \end{subfigure}\hfill
  \begin{subfigure}[b]{0.48\textwidth}
    \centering
    \includegraphics[width=\linewidth,height=6cm,keepaspectratio]{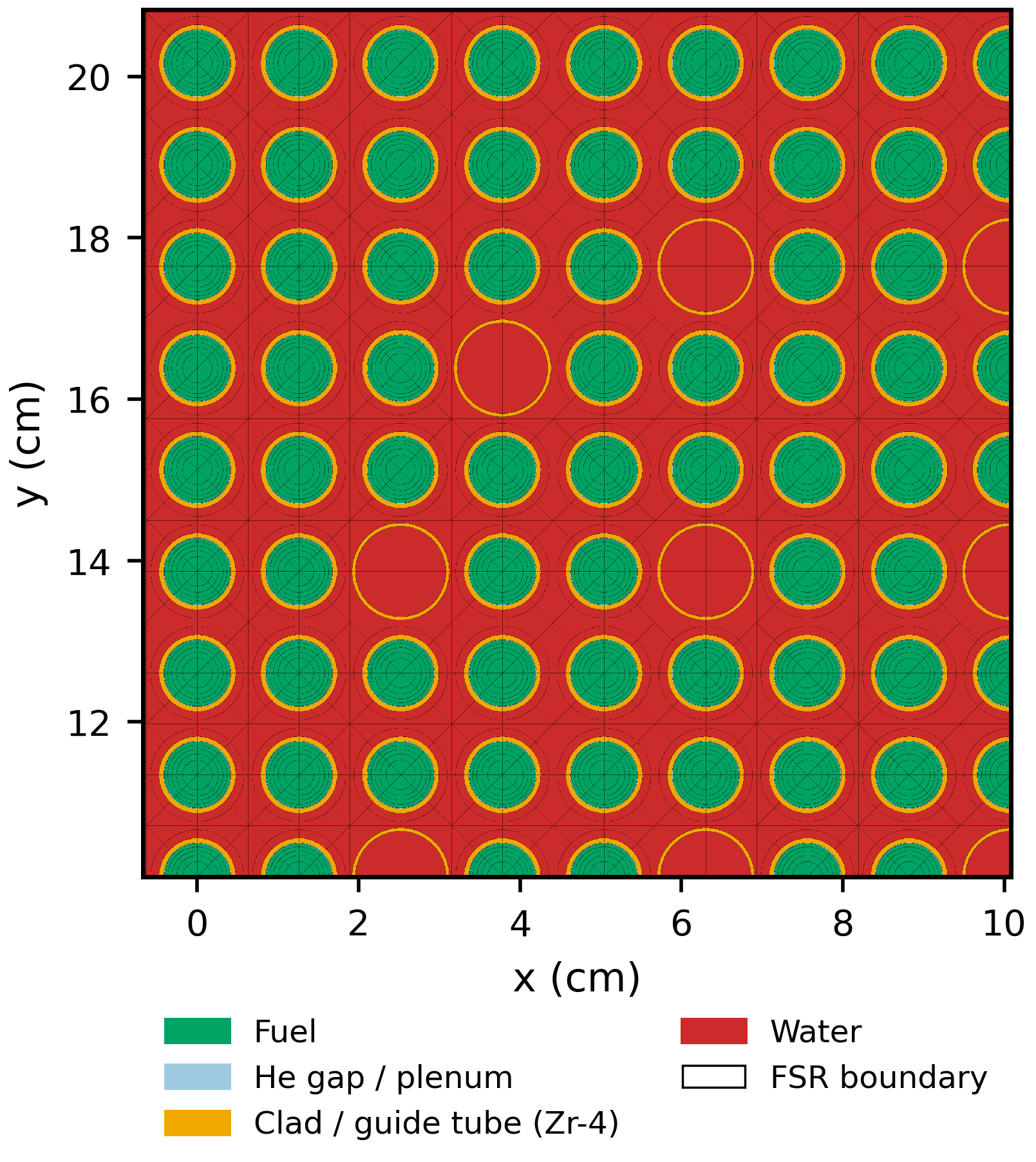}
    \caption{MOC FSRs, radial}
    \label{fig:17x17-moc-radial}
  \end{subfigure}
  \caption{Meshes of the $17\times17$ pin case: (a)--(b)~CFD cells, (c)--(d)~OpenMOC flat source regions.}
  \label{fig:17x17-meshes}
\end{figure}

Further verification is done, where a $17\times17$ PWR fuel assembly is modeled. Due to limits on computation memory, a quartered model was used instead, with reflective boundaries applied to all edges of the model. Verification against Problem 6 in the VERA core Physics Benchmark Progression Problems (\cite{godfrey2014vera}) is discussed in the later sections. Geometric and material input parameters are listed in Table.~\ref{tab:params_17x17} and Table.~\ref{tab:props_17x17} respectively, and geometric meshes are shown in Fig.~\ref{fig:17x17-meshes}. Coupling specific parameters are given in Table.~\ref{tab:coupling_17x17}. For mesh mapping, $17\times17\times100$ coupling bins were used, with MOC having 100 axial levels (1 level per bin) and CFD having 100 axial cells (1 cell per bin).

\begin{table}[pos=htbp]
  \caption{Parameters of the $17\times17$ fuel assembly model}
  \label{tab:params_17x17}
  \begin{tabular*}{\columnwidth}{@{\extracolsep{\fill}}ll@{}}
    \toprule
    Parameter & Value \\
    \midrule
    Fuel pellet radius (cm)              & 0.4096 \\
    Clad inner / outer radius (cm)       & 0.418 / 0.475 \\
    Guide tube inner / outer radius (cm) & 0.561 / 0.602 \\
    Rod pitch (cm)                       & 1.26 \\
    Assembly pitch (cm)                  & 21.50 \\
    Axial height (cm)                    & 365.76 \\
    Fuel                                 & UO$_2$ \\
    Fuel enrichment                      & 3.1\% \\
    Gap gas                              & Helium \\
    Clad / guide tube material           & Zircaloy-4 \\
    Coolant                              & Borated water \\
    Boron concentration (ppm)            & 1300 \\
    Inlet coolant temperature (K)        & 565 \\
    System pressure (MPa)                & 15.513 \\
    Mass flow rate (kg/s)                & 85.968 \\
    Power (MW)                           & 17.67 \\
    \bottomrule
  \end{tabular*}
\end{table}

\begin{table}[pos=H]
  \caption{Thermophysical properties of materials of $17\times17$ model. Thermal conductivities are taken from (\cite{fink1982uo2}) for UO$_2$ and (\cite{LuscherWalterG2010MPCC}}) for Zircaloy-4.
  \label{tab:props_17x17}
  \begin{tabular*}{\columnwidth}{@{\extracolsep{\fill}}llll@{}}
    \toprule
    Material   & Density    & Thermal conductivity   & Specific heat \\
               & (g/cm$^3$) & (W/m/K)                & (J/kg/K) \\
    \midrule
    UO$_2$     & 10.257     & Fink $k(T)\times0.906$ & 320 \\
    Zircaloy-4 & 6.56       & MATPRO $k(T)$          & 327.4 \\
    H$_2$O     & 0.7470     & 0.5806                 & 5286 \\
    \bottomrule
  \end{tabular*}
\end{table}

\begin{table*}[pos=tp]
  \caption{Parameters for the N-TH coupling of $17\times17$ model}
  \label{tab:coupling_17x17}
  \begin{tabular*}{\textwidth}{@{\extracolsep{\fill}}ll@{}}
    \toprule
    Parameter & Value \\
    \midrule
    \multicolumn{2}{@{}l}{\emph{MGXS generation (OpenMC)}} \\
    Particles per batch                     & 20{,}000 \\
    Inactive / total batches                & 50 / 200 \\
    Energy groups                           & 16 (CASMO) \\
    \addlinespace
    \multicolumn{2}{@{}l}{\emph{Neutronics (OpenMOC)}} \\
    Azimuthal / polar angles                & 16 / 6 \\
    Azimuthal / axial track spacing (cm)    & 0.1 / 1.0 \\
    Fuel rings / coolant rings / sectors    & 5 / 2 / 4 \\
    Axial levels                            & 100 \\
    CMFD relaxation factor                  & 0.7 \\
    Eigenvalue convergence criterion        & $10^{-5}$ \\
    \addlinespace
    \multicolumn{2}{@{}l}{\emph{Thermal-hydraulics (OpenFOAM)}} \\
    Number of cells in fuel region          & 1{,}689{,}600 \\
    Number of cells in cladding region      & 844{,}800 \\
    Number of cells in coolant region       & 3{,}514{,}000 \\
    Solver iterations per coupling window   & 1 \\
    Coupling windows per Picard iteration   & 22--600 \\
    \addlinespace
    \multicolumn{2}{@{}l}{\emph{Coupling}} \\
    Coupling bins                           & $17\times17\times100$ \\
    Data mapping                            & Nearest neighbour, consistent \\
    Power relaxation factor $\alpha$        & 0.5 \\
    Eigenvalue convergence criterion        & 5~pcm \\
    Power convergence criterion (RMS / max) & $10^{-3}$ / $10^{-2}$ \\
    \bottomrule
  \end{tabular*}
\end{table*}

\clearpage
\subsubsection{Results}
\label{sec:17x17_results}
\begin{figure}[pos=H]
  \centering
  \includegraphics[width=0.6\linewidth]{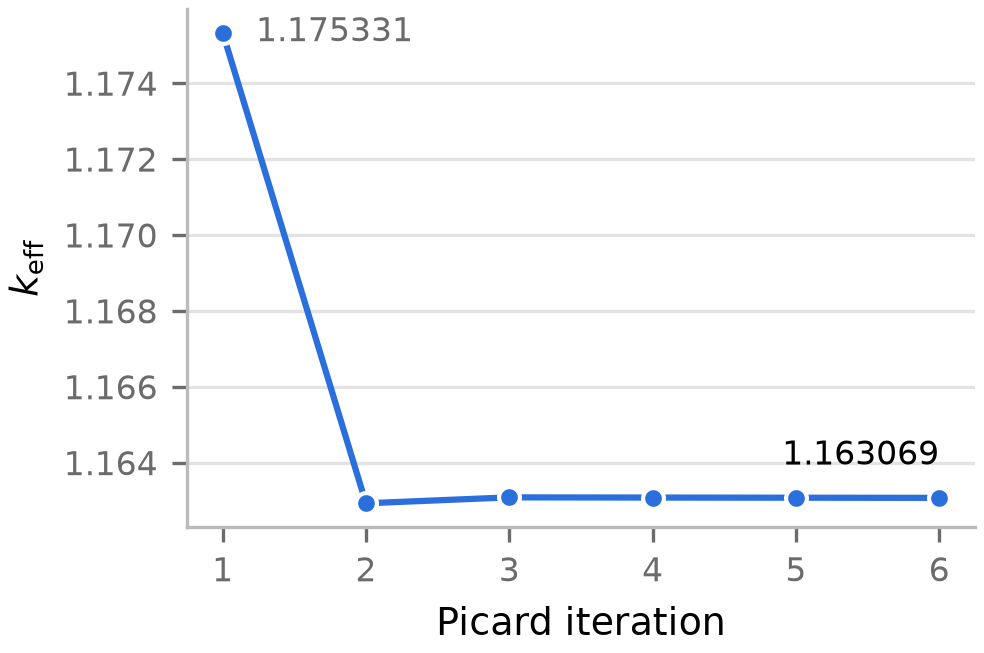}
  \caption{Variation of converged $k_\mathrm{eff}$ with iteration number for $17\times17$ case model}
  \label{fig:17x17_keff_v_iter}
\end{figure}

Fig.~\ref{fig:17x17_keff_v_iter} shows the change in $k_\mathrm{eff}$ over the iteration number. The solution approaches convergence within the second iteration, and is further fine tuned over the remaining iterations. The converged value of $k_\mathrm{eff}$ is 1.163069 (-54 pcm), in excellent agreement with the reference value of 1.16361 from VERA-CS in (~\cite{kelly2017}). Table~\ref{tab:17x17_comparison} summarizes the comparison of this study's calculated results with a few references' (\cite{kelly2017}, \cite{BullerwellLanceEdward2020AMCS}, \cite{SchmidtRodney2015Aafc}) stated VERA solutions.

\begin{figure}[pos=H]
  \centering
  \includegraphics[width=0.6\linewidth]{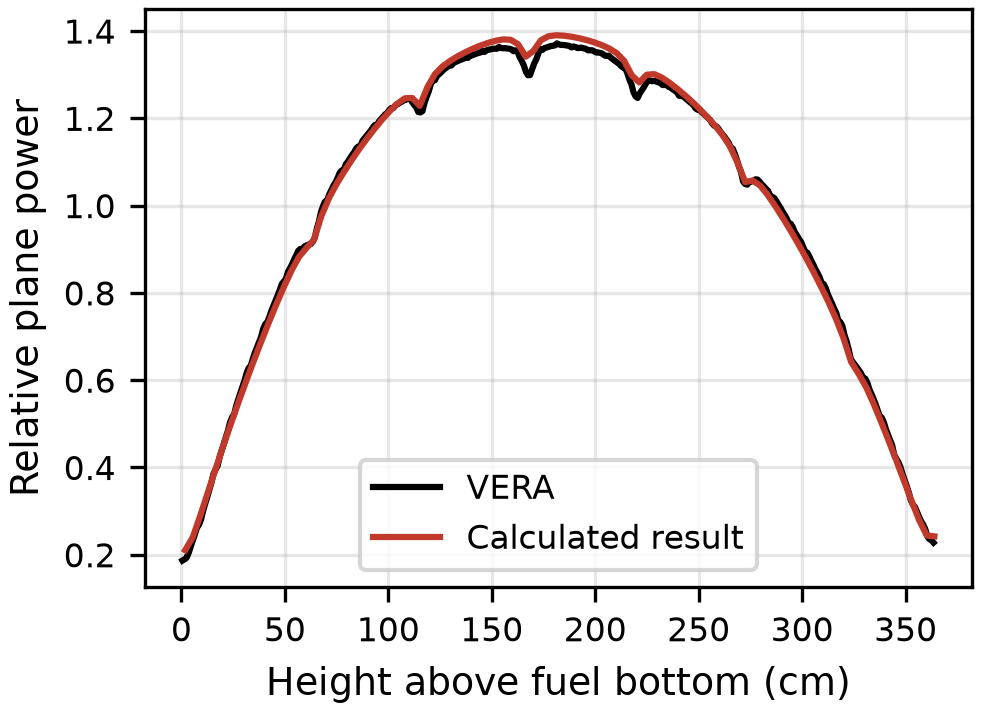}
  \caption{Comparison of axial profile against VERA for $17\times17$ case model}
  \label{fig:17x17_axial_vs_vera}
\end{figure}

\begin{figure}[pos=htbp]
  \begin{subfigure}[b]{0.48\textwidth}
    \centering
    \includegraphics[width=\linewidth]{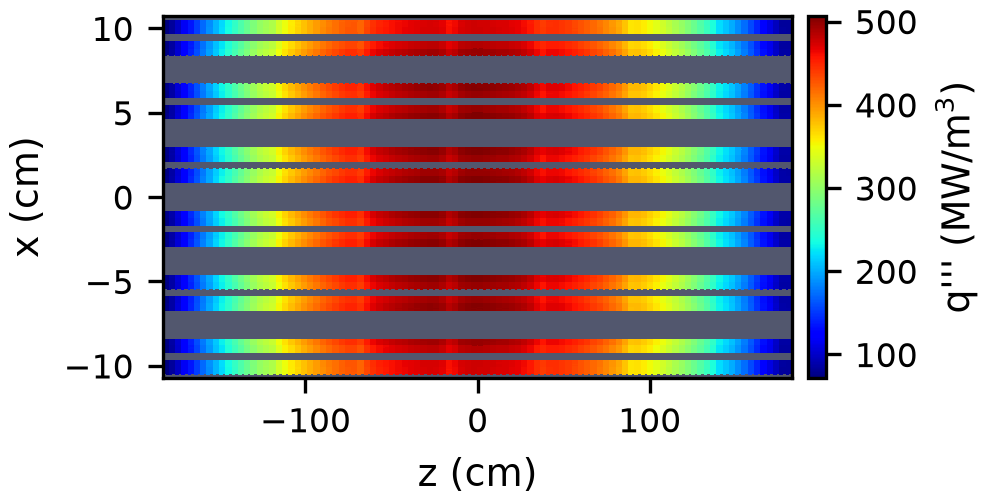}
    \caption{Calculated axial power at centerline}
    \label{fig:17x17_calc_power_axial_centerline}
  \end{subfigure}\hfill
  \begin{subfigure}[b]{0.48\textwidth}
    \centering
    \includegraphics[width=\linewidth]{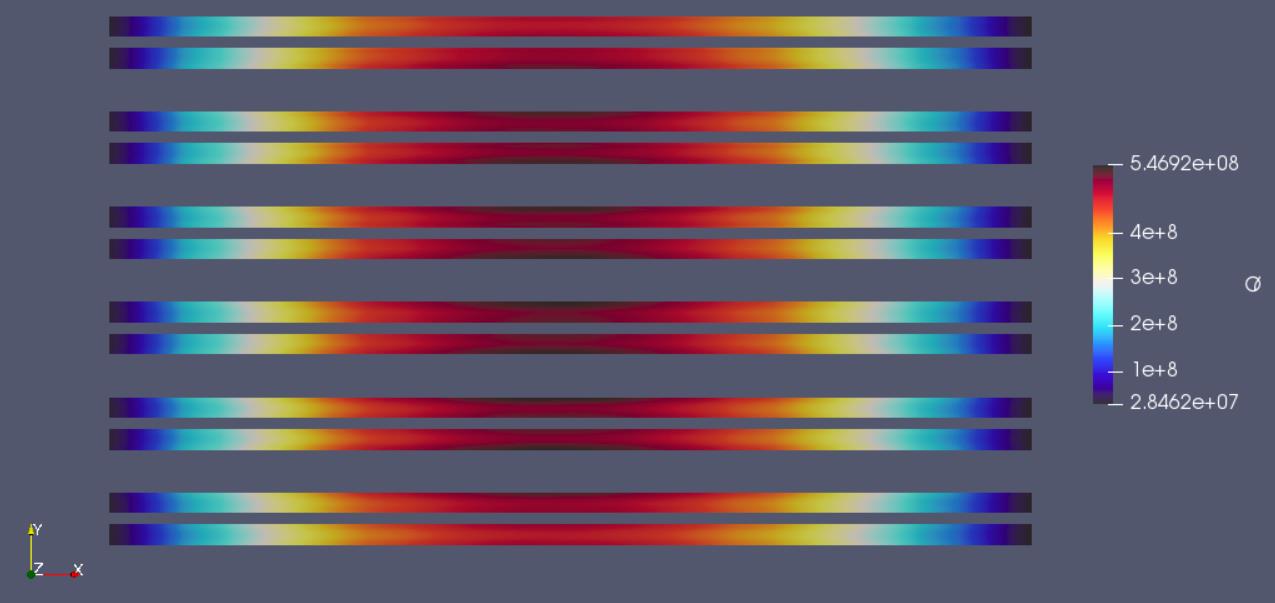}
    \caption{Reference axial power at centerline}
    \label{fig:17x17_ref_power_axial_centerline}
  \end{subfigure}
  \caption{Axial power at centerline of $17\times17$ model: (a)Calculated result, (b)Reference result}
  \label{fig:17x17_power_axial_centerline}
\end{figure}

\begin{figure}[pos=htbp]
  \begin{subfigure}[b]{0.48\textwidth}
    \centering
    \includegraphics[width=\linewidth]{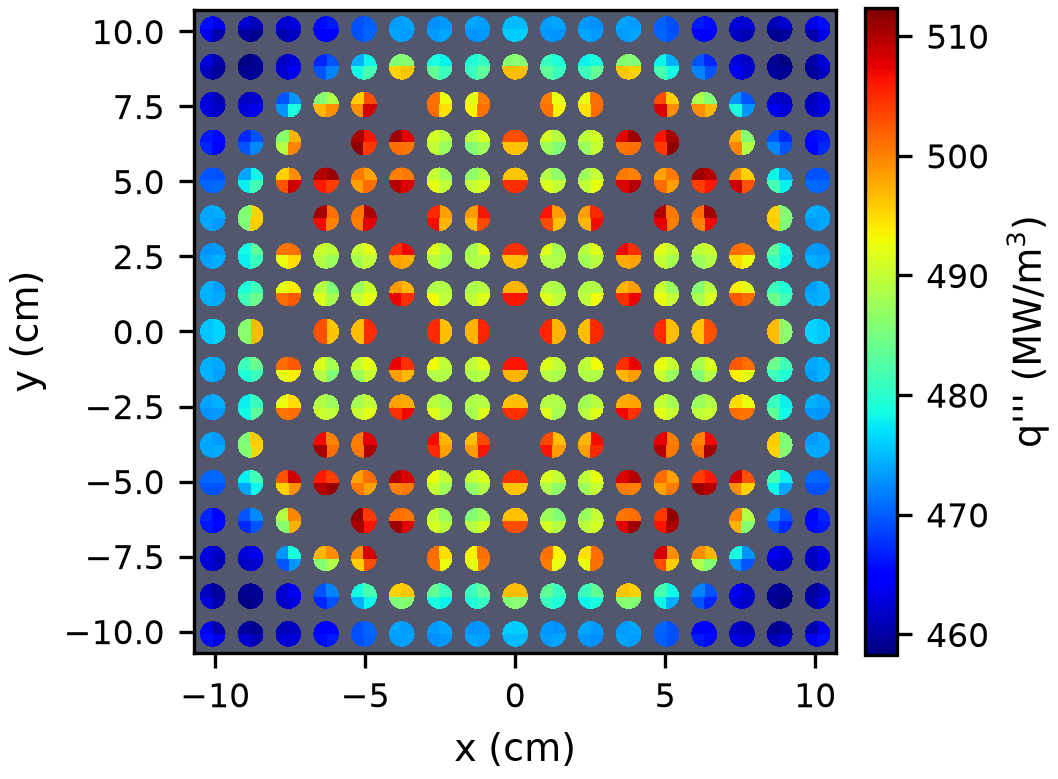}
    \caption{Calculated radial power at midpoint}
    \label{fig:17x17_calc_power_radial_mid}
  \end{subfigure}\hfill
  \begin{subfigure}[b]{0.48\textwidth}
    \centering
    \includegraphics[width=\linewidth]{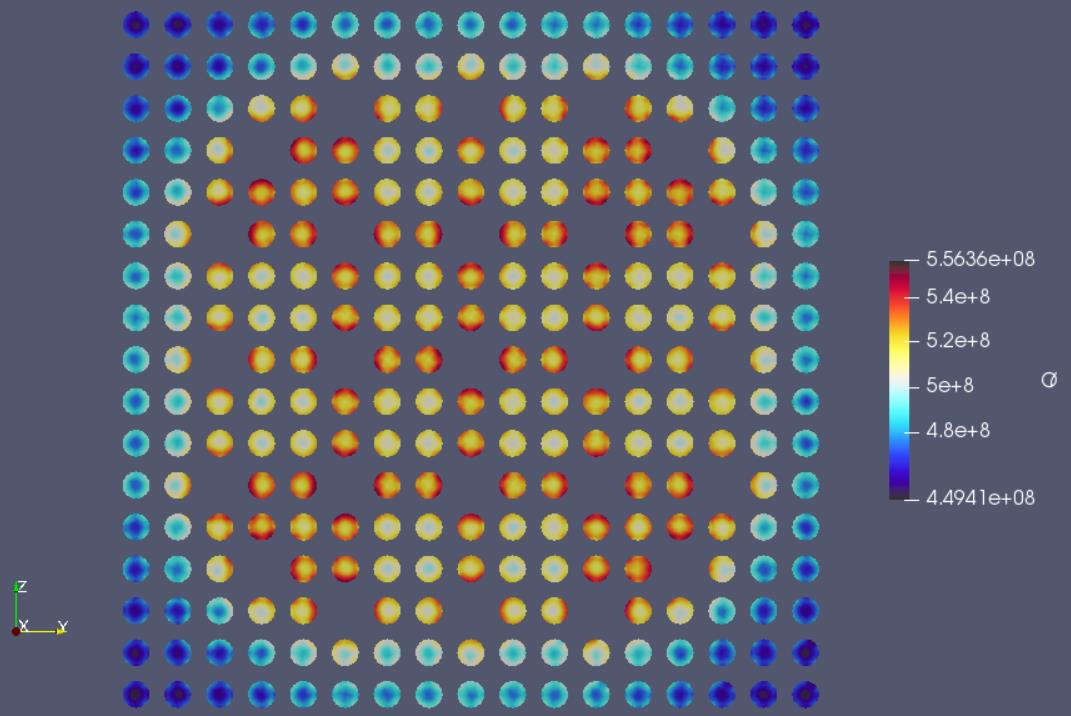}
    \caption{Reference radial power at midpoint}
    \label{fig:17x17_ref_power_radial_mid}
  \end{subfigure}
  \caption{Radial power at midpoint of $17\times17$ model: (a)Calculated result, (b)Reference result}
  \label{fig:17x17_power_radial_mid}
\end{figure}

\begin{figure}[pos=htbp]
  \begin{subfigure}[b]{0.48\textwidth}
    \centering
    \includegraphics[width=\linewidth]{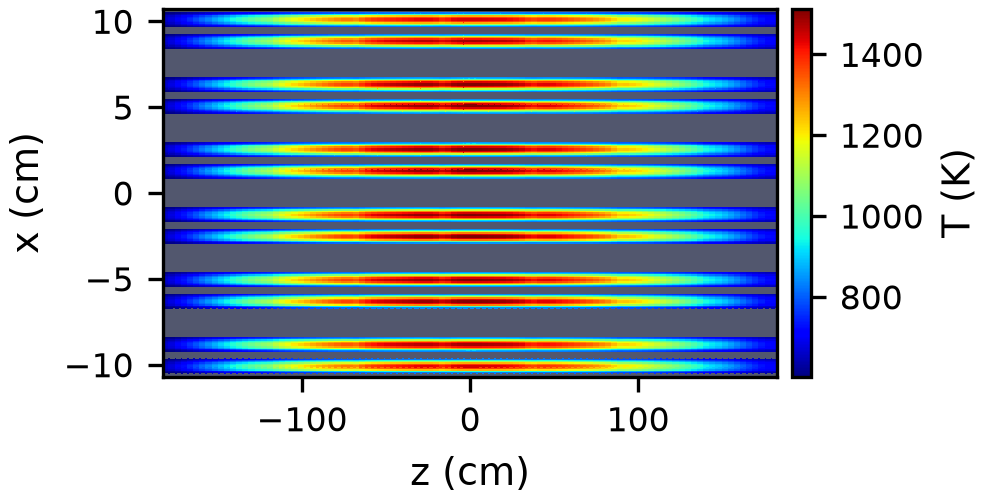}
    \caption{Calculated axial fuel temperature}
    \label{fig:17x17_calc_fuel_T_axial}
  \end{subfigure}\hfill
  \begin{subfigure}[b]{0.48\textwidth}
    \centering
    \includegraphics[width=\linewidth]{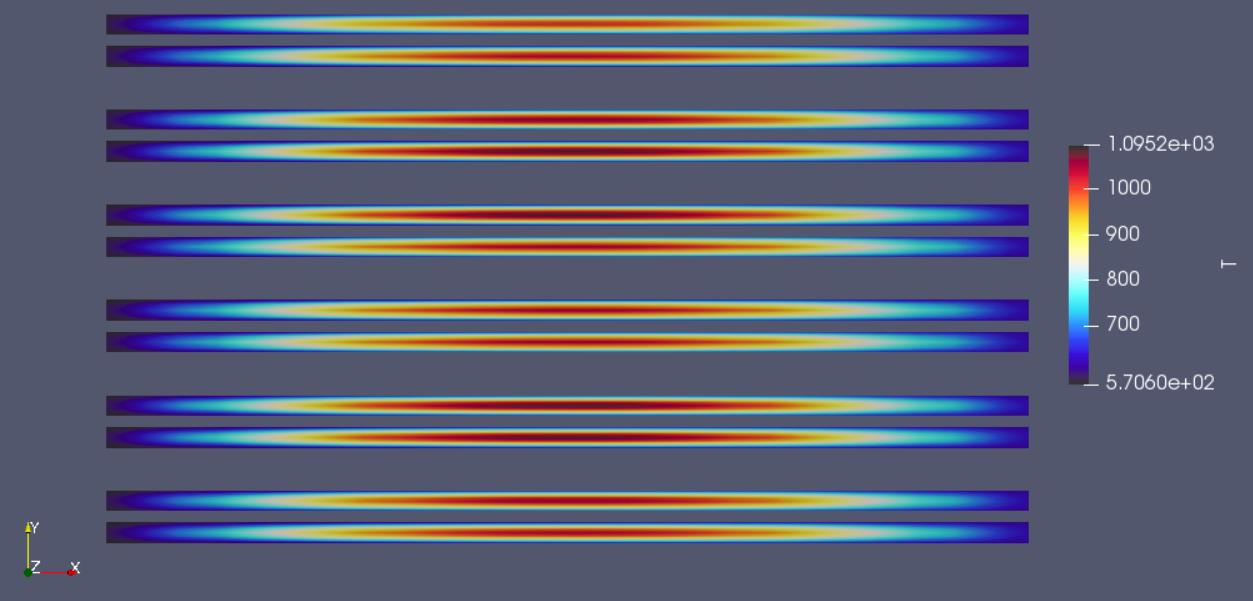}
    \caption{Reference axial fuel temperature}
    \label{fig:17x17_ref_fuel_T_axial}
  \end{subfigure}
  \caption{Axial fuel temperature of $17\times17$ model: (a)Calculated result, (b)Reference result}
  \label{fig:17x17_fuel_T_axial}
\end{figure}

\begin{figure}[pos=htbp]
  \begin{subfigure}[b]{0.48\textwidth}
    \centering
    \includegraphics[width=\linewidth]{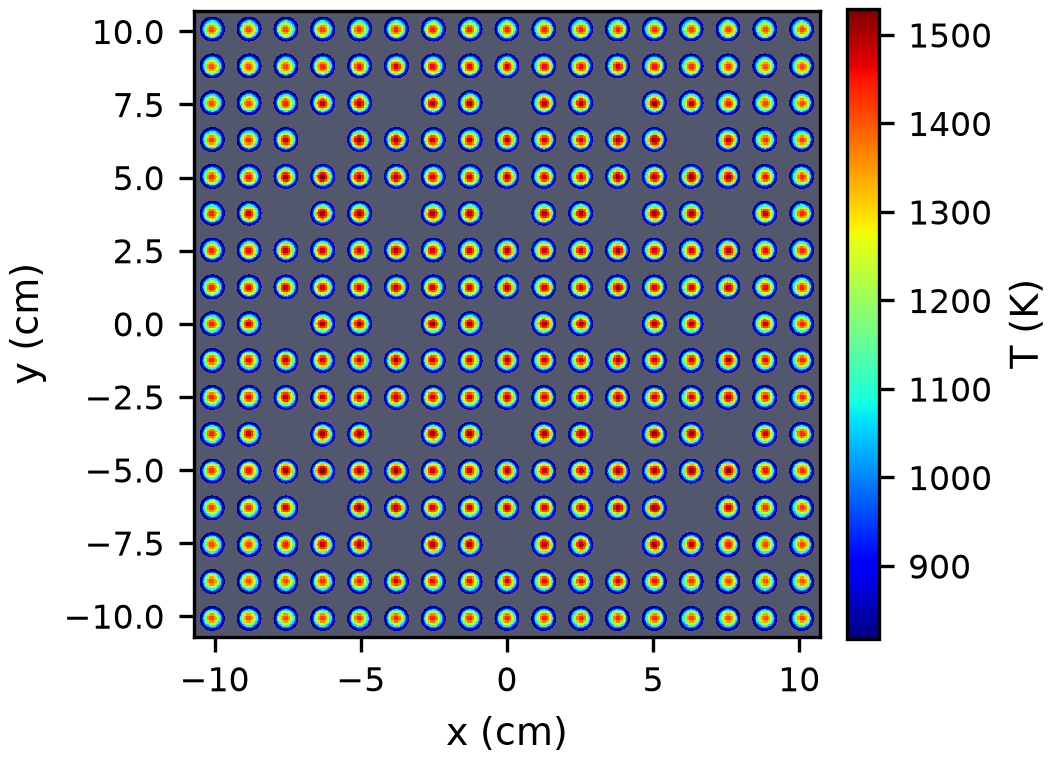}
    \caption{Calculated radial fuel temperature}
    \label{fig:17x17_calc_fuel_T_radial}
  \end{subfigure}\hfill
  \begin{subfigure}[b]{0.48\textwidth}
    \centering
    \includegraphics[width=\linewidth]{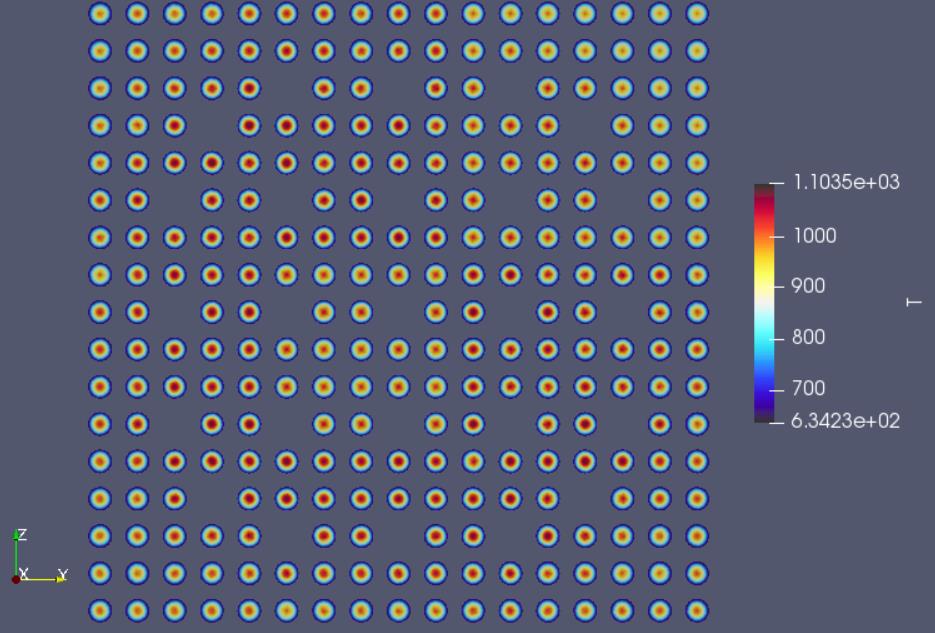}
    \caption{Reference radial fuel temperature}
    \label{fig:17x17_ref_fuel_temp_radial}
  \end{subfigure}
  \caption{Radial fuel temperature at midpoint of $17\times17$ model: (a)Calculated result, (b)Reference result}
  \label{fig:17x17_fuel_temp_radial}
\end{figure}

\begin{figure}[pos=htbp]
  \begin{subfigure}[b]{0.48\textwidth}
    \centering
    \includegraphics[width=\linewidth]{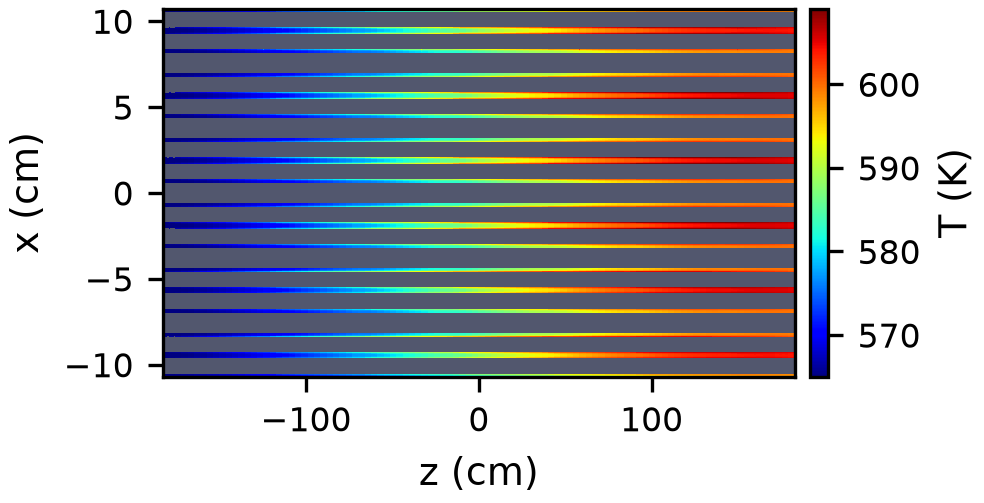}
    \caption{Calculated axial moderator temperature at centerline}
    \label{fig:17x17_calc_water_T_axial}
  \end{subfigure}\hfill
  \begin{subfigure}[b]{0.48\textwidth}
    \centering
    \includegraphics[width=\linewidth]{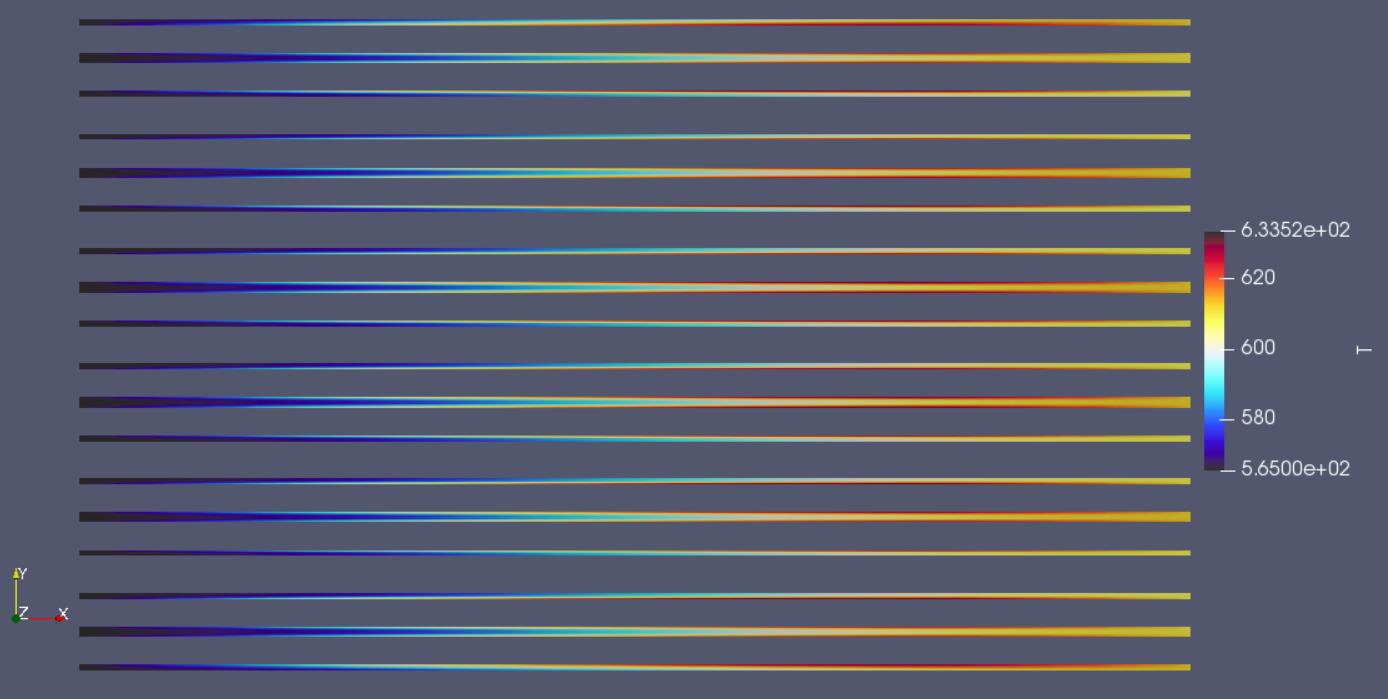}
    \caption{Reference axial moderator temperature at centerline}
    \label{fig:17x17_ref_water_T_axial}
  \end{subfigure}
  \caption{Axial moderator temperature at centerline of $17\times17$ model: (a)Calculated result, (b)Reference result}
  \label{fig:17x17_water_T_axial}
\end{figure}

\begin{figure}[pos=htbp]
  \begin{subfigure}[b]{0.48\textwidth}
    \centering
    \includegraphics[width=\linewidth]{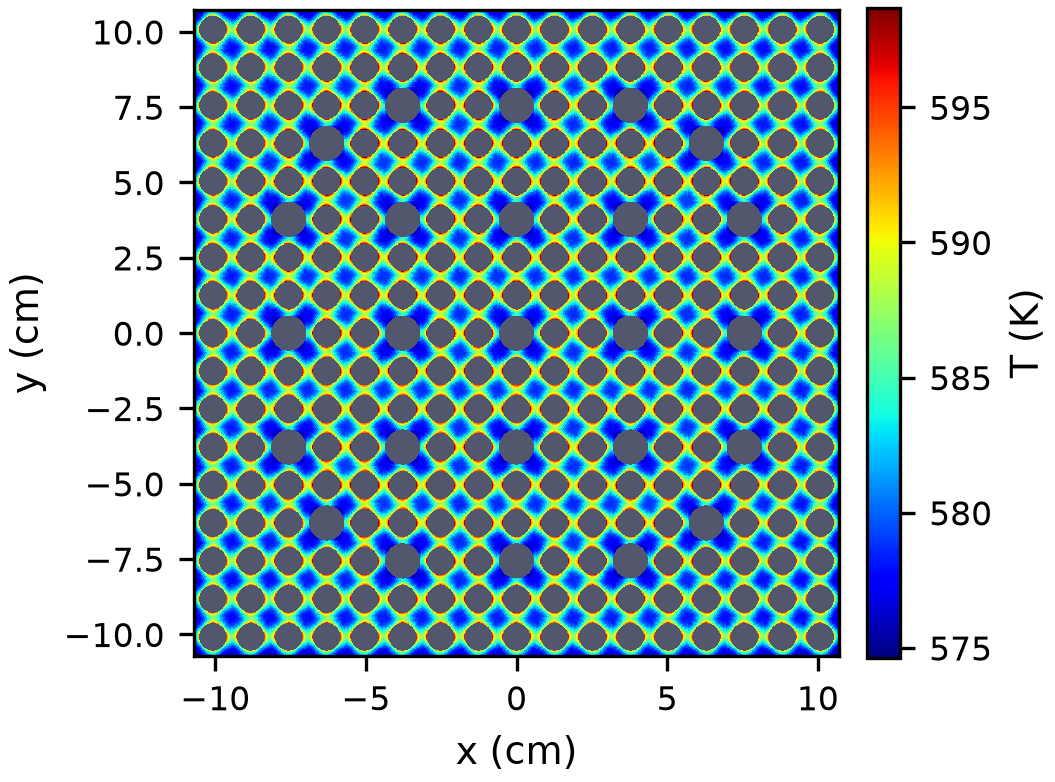}
    \caption{Calculated radial moderator temperature}
    \label{fig:17x17_calc_water_T_radial}
  \end{subfigure}\hfill
  \begin{subfigure}[b]{0.48\textwidth}
    \centering
    \includegraphics[width=\linewidth]{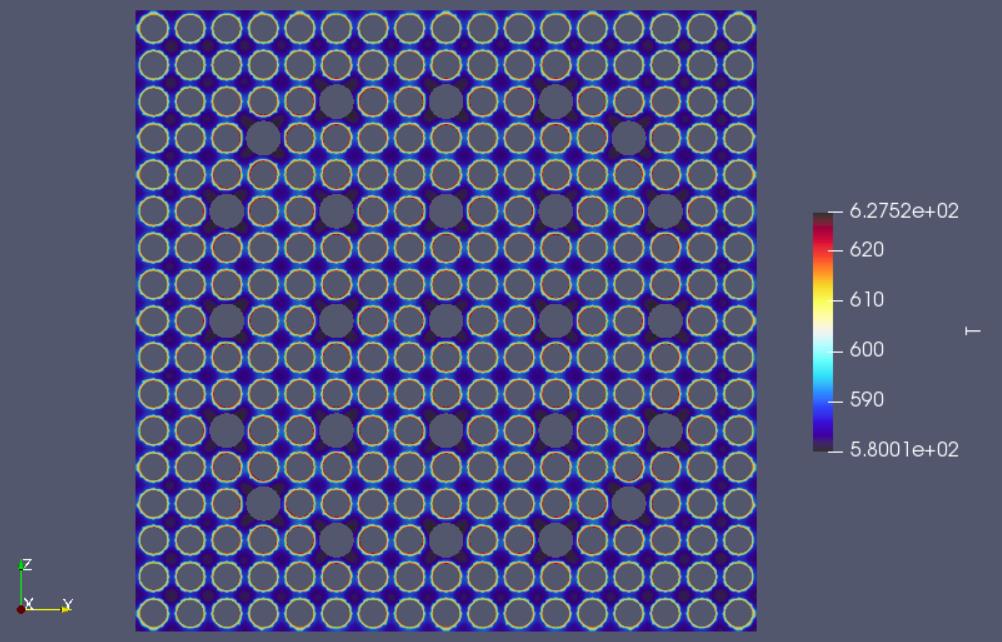}
    \caption{Reference radial moderator temperature}
    \label{fig:17x17_ref_water_T_radial}
  \end{subfigure}
  \caption{Radial moderator temperature at midpoint of $17\times17$ model: (a)Calculated result, (b)Reference result}
  \label{fig:17x17_water_T_radial}
\end{figure}

\begin{figure}[pos=htbp]
  \begin{subfigure}[b]{0.48\textwidth}
    \centering
    \includegraphics[width=\linewidth]{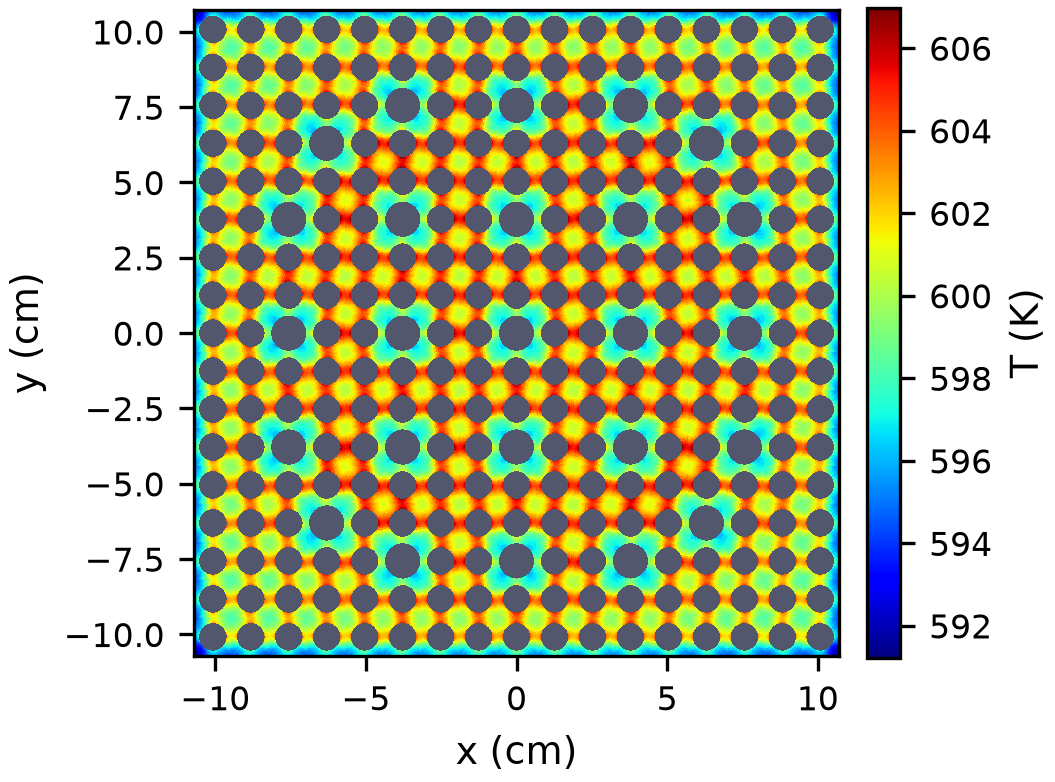}
    \caption{Calculated radial moderator temperature at outlet}
    \label{fig:17x17_calc_water_T_radial_out}
  \end{subfigure}\hfill
  \begin{subfigure}[b]{0.48\textwidth}
    \centering
    \includegraphics[width=\linewidth]{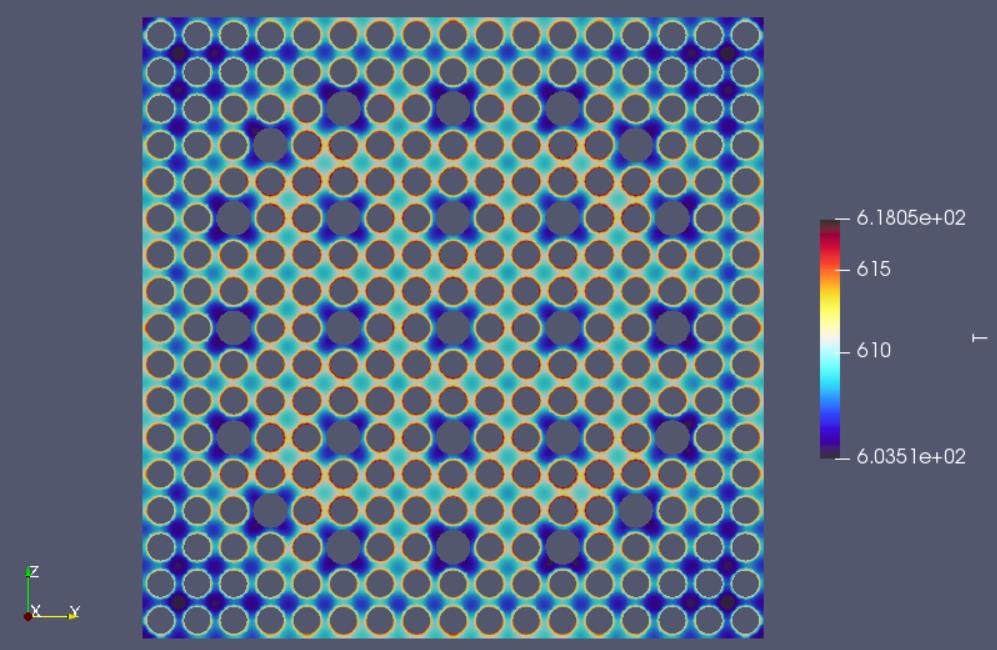}
    \caption{Reference radial moderator temperature at outlet}
    \label{fig:17x17_ref_water_T_radial_out}
  \end{subfigure}
  \caption{Radial moderator temperature at outlet of $17\times17$ model: (a)Calculated result, (b)Reference result}
  \label{fig:17x17_water_T_radial_out}
\end{figure}

Figures shown will be comparisons of this study's calculated results with the reference results of (~\cite{BullerwellLanceEdward2020AMCS}). Fig.~\ref{fig:17x17_power_axial_centerline} and Fig.~\ref{fig:17x17_power_radial_mid} compares the calculated power distribution results; Fig.~\ref{fig:17x17_fuel_T_axial} and Fig.~\ref{fig:17x17_fuel_temp_radial} compares the fuel temperature distribution; Fig.~\ref{fig:17x17_water_T_axial}, Fig.~\ref{fig:17x17_water_T_radial} and Fig.~\ref{fig:17x17_water_T_radial_out} compares the moderator temperature distribution. As shown in Table~\ref{tab:17x17_comparison}, the results of power distribution and moderator temperature distribution are in good agreement with reference VERA solutions. Referring to Fig.~\ref{fig:17x17_axial_vs_vera}, displaying the shape of the relative power axial distribution, it can be seen that the calculated results closely match the shape produced by the VERA solution. This reaffirms that the Physics has been applied accurately in the coupling framework. Fig.~\ref{fig:17x17_power_axial_centerline} and Fig.~\ref{fig:17x17_fuel_T_axial} both show that this study's calculated results are skewed slightly towards negative z. This can be explained since the coolant flow enters from negative z and flows to positive z. As such the coolant near the inlet (negative z) is colder and denser than at the outlet (positive z). As discussed in Section~\ref{sec:N-TH_feedback}, the denser coolant moderates neutrons more effectively, increasing the fission rate in the lower half of the assembly. The opposite effect applies to the upper half of the assembly. 

\begin{table*}[pos=t]
  \caption{Comparison of the $17\times17$ assembly results (VERA Problem~6, 1300~ppm) with reference solutions}
  \label{tab:17x17_comparison}
  \begin{tabular*}{\textwidth}{@{\extracolsep{\fill}}lcclc@{}}
    \toprule
    Quantity & Present work & Reference & Source & Difference \\
    \midrule
    \multicolumn{5}{@{}l}{\emph{Neutronics}} \\
    $k_\mathrm{eff}$                     & 1.163069 & 1.16361  & VERA-CS \citep{kelly2017}                          & $-54$~pcm \\
    Axial peaking factor                 & 1.387    & 1.45064  & VERA, via \citet[Table~4]{BullerwellLanceEdward2020AMCS} & $-4.39$\% \\
    Max.\ pin power, pin (4,5)           & 1.046    & 1.0517   & VERA-CS \citep{kelly2017}                          & $-0.54$\% \\
    Min.\ pin power, pin (7,7)           & 0.946    & 0.940    & VERA-CS \citep{kelly2017}                          & $+0.64$\% \\
    Corner pin power, pin (8,8)          & 0.9555   & 0.9487   & VERA-CS \citep{kelly2017}                          & $+0.72$\% \\
    \midrule
    \multicolumn{5}{@{}l}{\emph{Thermal-hydraulics}} \\
    Avg.\ coolant density (g/cm$^3$)     & 0.700274 & 0.698785 & VERA, via \citet[Table~4]{BullerwellLanceEdward2020AMCS} & $+0.21$\% \\
    Coolant outlet temperature (K)       & 600.5    & 601      & VERA Insilico/CTF \citep{SchmidtRodney2015Aafc}$^{a}$ & $-0.5$~K \\
    Subchannel exit $T$, mean (K)        & 600.15   & 599.75   & VERA-CS \citep{kelly2017}                          & $+0.4$~K \\
    Subchannel exit $T$, min (K)         & 593.75   & 595.05   & VERA-CS \citep{kelly2017}                          & $-1.3$~K \\
    Subchannel exit $T$, max (K)         & 603.15   & 601.65   & VERA-CS \citep{kelly2017}                          & $+1.5$~K \\
    \midrule
    \multicolumn{5}{@{}l}{\emph{Fuel temperature}} \\
    Peak plane-avg.\ fuel $T$ (K)        & 1127.2   & \multirow{2}{*}{1117.1} & \multirow{2}{*}{VERA, via \citet[Table~4]{BullerwellLanceEdward2020AMCS}$^{*}$} & $+10.1$~K \\
    Peak pin-avg.\ fuel $T$ (K)          & 1161.3   &                         &                                                                                & $+44.2$~K \\
    \bottomrule
    \multicolumn{5}{@{}p{\textwidth}@{}}{\footnotesize
      $^{a}$\,Nominal case at 600~ppm boron, not 1300~ppm.
      $^{*}$\,Reported as unspecified max fuel temperature.}
  \end{tabular*}
\end{table*}

One of the fundamental motivations for this study is the total time taken to run the coupled simulation at scale. The time breakdown of each solver, excluding overhead wall time, is listed down in Table~\ref{tab:17x17_runtime}. The run time of this study's coupled simulation is 39.7 core hours. Core allocation influences the wall-clock time, with the thermal-hydraulic solver (Water) being the most time consuming, but it is not the main reason for the low cost. The low computational cost stems from the numerical methods (deterministic OpenMOC instead of stochastic OpenMC) and the modeling simplifications: Due to memory limitations, this study opted for a quartered $17\times17$ single fuel assembly. This study also uses the 16-group CASMO for the multi-group cross sections, which aids in accelerating the simulation.

\begin{table*}[pos=tp]
  \caption{Core allocation and computational cost of the $17\times17$ coupled
    calculation. The run converged after 6 Picard iterations
    (581 coupling windows).}
  \label{tab:17x17_runtime}
  \begin{tabular*}{\textwidth}{@{\extracolsep{\fill}}lcccc@{}}
    \toprule
    Participant & Solver & Cores & Time (h) & Core-hours \\
    \midrule
    MGXS generation$^{b}$ & OpenMC         & 8$^{c}$ & 3.04 & 24.3 \\
    \midrule
    Fuel  & \texttt{laplacianFoam}     & 1       & 0.76 & 0.8 \\
    Clad  & \texttt{laplacianFoam}     & 1       & 0.54 & 0.5 \\
    Water & \texttt{buoyantSimpleFoam} & 3       & 3.53 & 10.6 \\
    MOC   & OpenMOC                    & 3$^{a}$ & 1.41 & 4.2 \\
    Coupled run (wall clock)           &         & 8 & 4.96 & 39.7 \\
    \midrule
    Overall (wall clock)$^{d}$         &         & 8 & 8.00 & 64.0 \\
    \bottomrule
    \multicolumn{5}{@{}p{\textwidth}@{}}{\footnotesize
      $^{a}$\,OpenMP threads.
      $^{b}$\,Reference library, fuel and water temperature sweeps, fuel tables at two
      coolant densities, spacer-grid and axial-reflector libraries, run once before the
      coupled calculation.
      $^{c}$\,Physical cores (OpenMP threads, OpenMC default).
      $^{d}$\,MGXS generation plus coupled run, run one after the other.}
  \end{tabular*}
\end{table*}

\newpage
\section{Conclusions}
\label{sec:conclusion}
This paper developed and validated a computationally efficient neutronics–thermal-hydraulics (N–TH) coupling framework integrating the deterministic neutron transport solver OpenMOC with the computational fluid dynamics solver OpenFOAM through the multiphysics coupling library preCICE. 
The framework addresses the current high computational cost of conventional Monte Carlo-based coupled simulations by employing the method of characteristics (MOC) for repeated neutronics calculations, while using OpenMC for the initial generation of multigroup cross-section libraries.

Our proposed framework was validated using a simplified 2 × 2 fuel lattice and a quartered 17 × 17 pressurized water reactor fuel assembly. 
The results demonstrated reasonable agreement with established reference solutions. 
For the 17 × 17 assembly, the calculated effective multiplication factor differed from the VERA reference by only 54 pcm, while deviations in selected pin-wise power factors remained below 1 percent. The coolant outlet temperature also showed close agreement with the reference solution.

Significant computational efficiency was demonstrated. For the 2 × 2 configuration, the coupled simulation required approximately 2.7 hours including initial cross-section generation, compared with 17.6 hours reported for a previous OpenMC–OpenFOAM calculation. 
For the quartered 17 × 17 assembly, the coupled simulation is only 5 hours or 8 hours including cross-section generation. 
These results highlight the potential advantages of deterministic neutronics for repeated coupled calculations, particularly when precomputed cross-section libraries can be reused.

The present implementation is limited to steady-state, single-phase conditions and has been demonstrated only up to a quartered fuel assembly. Future work will focus on transient coupling, larger reactor configurations, numerical convergence studies and improved computational scalability.
Ultimately, the proposed framework offers a promising foundation for efficient reactor multiphysics simulations, parametric investigations and potential nuclear reactor digital-twin applications.



\section*{CRediT authorship contribution statement}
\textbf{Seth S. W. Choo:} Conceptualization, Methodology, Software, Validation, Writing -- original draft
\textbf{L. K. Ang:} Supervision, Writing -- review \& editing
\textbf{Zhe Chuan Tan:} Conceptualization, Supervision, Resources, Writing -- review \& editing

\section*{Declaration of competing interest}
The authors declare that they have no known competing financial interests or personal relationships that could have appeared to influence the work reported in this paper.

\section*{Data availability}
Data will be made available on request.

\section*{Acknowledgements}
This research did not receive any specific grant from funding agencies in the public, commercial, or not-for-profit sectors.

\FloatBarrier
\bibliographystyle{cas-model2-names}   
\bibliography{references}

@article{ZHANG2026106495,
title = {Development of a general neutronics–thermal hydraulics coupling framework based on partitioned coupling},
journal = {Progress in Nuclear Energy},
volume = {199},
pages = {106495},
year = {2026},
issn = {0149-1970},
doi = {https://doi.org/10.1016/j.pnucene.2026.106495},
url = {https://www.sciencedirect.com/science/article/pii/S014919702600260X},
author = {Dacai Zhang and Huifu Wang and Zeguang Li and Zilin Su and Xuefeng Tan and Pengfei Shen}
}

@article{BOYD201443,
title = {The OpenMOC method of characteristics neutral particle transport code},
journal = {Annals of Nuclear Energy},
volume = {68},
pages = {43-52},
year = {2014},
issn = {0306-4549},
doi = {https://doi.org/10.1016/j.anucene.2013.12.012},
url = {https://www.sciencedirect.com/science/article/pii/S0306454913006634},
author = {William Boyd and Samuel Shaner and Lulu Li and Benoit Forget and Kord Smith}
}

@article{zhang2021mcfluent,
author = {Zhang, Qingyang and Peng, Tianji and Zhang, Guangchun and Liu, Jie and Guo, Xiaowei and Gong, Chunye and Yang, Bo and Fan, Xukai},
title = {An Efficient Scheme for Coupling OpenMC and FLUENT with Adaptive Load Balancing},
journal = {Science and Technology of Nuclear Installations},
volume = {2021},
number = {1},
pages = {5549602},
doi = {https://doi.org/10.1155/2021/5549602},
url = {https://onlinelibrary.wiley.com/doi/abs/10.1155/2021/5549602},
eprint = {https://onlinelibrary.wiley.com/doi/pdf/10.1155/2021/5549602},
year = {2021}
}

@article{LOI2024113480,
title = {OFELIA: An OpenMC-FEniCSx coupling for neutronic calculation with temperature feedback},
journal = {Nuclear Engineering and Design},
volume = {428},
pages = {113480},
year = {2024},
issn = {0029-5493},
doi = {https://doi.org/10.1016/j.nucengdes.2024.113480},
url = {https://www.sciencedirect.com/science/article/pii/S0029549324005806},
author = {Lorenzo Loi and Stefano Riva and Carolina Introini and Francesca Giacobbo and Xiang Wang and Antonio Cammi}
}

@article{ZHAO2025109724,
title = {Neutronic and thermal-hydraulics coupling analyses on prismatic TRISO-fueled gas-cooled reactors using Monte Carlo and computational fluid dynamics method},
journal = {Computer Physics Communications},
volume = {315},
pages = {109724},
year = {2025},
issn = {0010-4655},
doi = {https://doi.org/10.1016/j.cpc.2025.109724},
url = {https://www.sciencedirect.com/science/article/pii/S0010465525002267},
author = {Xingyu Zhao and Guodong Liu and Shanfang Huang and Kan Wang and Qiaoyan Chen and Guoming Liu}
}

@article{ROMANO201590,
title = {OpenMC: A state-of-the-art Monte Carlo code for research and development},
journal = {Annals of Nuclear Energy},
volume = {82},
pages = {90-97},
year = {2015},
note = {Joint International Conference on Supercomputing in Nuclear Applications and Monte Carlo 2013, SNA + MC 2013. Pluri- and Trans-disciplinarity, Towards New Modeling and Numerical Simulation Paradigms},
issn = {0306-4549},
doi = {https://doi.org/10.1016/j.anucene.2014.07.048},
url = {https://www.sciencedirect.com/science/article/pii/S030645491400379X},
author = {Paul K. Romano and Nicholas E. Horelik and Bryan R. Herman and Adam G. Nelson and Benoit Forget and Kord Smith}
}

@article{chourdakis2023openfoam,
  author  = {Chourdakis, Gerasimos and Schneider, David and Uekermann, Benjamin},
  title   = {{OpenFOAM-preCICE}: Coupling {OpenFOAM} with External Solvers for Multi-Physics Simulations},
  journal = {OpenFOAM Journal},
  volume  = {3},
  pages   = {1--25},
  year    = {2023},
  doi     = {10.51560/ofj.v3.88}
}

@article{FengKuaiyuan2025DoaM,
author = {Feng, Kuaiyuan and Song, Qufei and Wang, Yihu and Guo, Hui and Cong, Tenglong and Xiao, Yao and Gu, Hanyang},
copyright = {2025 Korean Nuclear Society},
issn = {1738-5733},
journal = {Nuclear engineering and technology},
language = {eng},
number = {9},
pages = {103645-14},
publisher = {Elsevier B.V},
title = {Development of a MOC-based multiphysics coupling method for analysis of FCM-fueled micro gas-cooled reactor},
volume = {57},
year = {2025},
}

@article{StauffNicolasE.2024HMMo,
author = {Stauff, Nicolas E. and Miao, Yinbin and Cao, Yan and Mo, Kun and Abdelhameed, Ahmed Amin E. and Ibarra, Lander and Matthews, Christopher and Shemon, Emily R.},
address = {United States},
issn = {0029-5639},
journal = {Nuclear science and engineering},
language = {eng},
organization = {Argonne National Laboratory (ANL), Argonne, IL (United States)},
pages = {1-17},
publisher = {Taylor \& Francis},
title = {High-Fidelity Multiphysics Modeling of a Heat Pipe Microreactor Using BlueCrab},
year = {2024},
}

@article{MatthewsChristopher2021CMSo,
author = {Matthews, Christopher and Laboure, Vincent and DeHart, Mark and Hansel, Joshua and Andrs, David and Wang, Yaqi and Ortensi, Javier and Martineau, Richard C.},
address = {La Grange Park},
copyright = {This work was authored as part of the Contributor's official duties as an Employee of the United States Government and is therefore a work of the United States Government. In accordance with 17 U.S.C. 105, no copyright protection is available for such works under U.S. Law.},
issn = {0029-5450},
journal = {Nuclear technology},
language = {eng},
number = {7},
organization = {Los Alamos National Laboratory (LANL), Los Alamos, NM (United States)},
pages = {1142-1162},
publisher = {Taylor \& Francis},
title = {Coupled Multiphysics Simulations of Heat Pipe Microreactors Using DireWolf},
volume = {207},
year = {2021},
}

@article{FengWenpei2021Doam,
author = {Feng, Wenpei and Jiang, Rong and Chen, Hongli and Qiang, Shenglong and Gong, Zhaohu and Li, Zhigang and Pan, Junjie and Zhao, Wenbo and Zhang, Xilin and Luo, Xiao and Zhang, Xue},
copyright = {2021 Elsevier Ltd},
issn = {0306-4549},
journal = {Annals of nuclear energy},
language = {eng},
pages = {108178-},
publisher = {Elsevier Ltd},
title = {Development of a multi-physics coupling system based on ICoCo interface and its validation on NEA-OECD core transient benchmark},
volume = {156},
year = {2021},
}

@article{Gunow2019BEAVRS,
  author  = {Gunow, Geoffrey and Forget, Benoit and Smith, Kord},
  title   = {Full core {3D} simulation of the {BEAVRS} benchmark with {OpenMOC}},
  journal = {Annals of Nuclear Energy},
  year    = {2019},
  doi     = {10.1016/j.anucene.2019.05.050}
}

@inproceedings{Gunow2017MC,
  author    = {Gunow, Geoffrey and Shaner, Samuel and Boyd, William and Forget, Benoit and Smith, Kord},
  title     = {Accuracy and Performance of {3D} {MOC} for Full-Core {PWR} Problems},
  booktitle = {Proc. M\&C 2017 -- International Conference on Mathematics \& Computational Methods Applied to Nuclear Science \& Engineering},
  address   = {Jeju, Korea},
  month     = apr,
  year      = {2017},
  note      = {April 16--20, 2017, on USB}
}

@article{WangBo2025Raao,
author = {Wang, Bo and Xie, Zeyi and Huang, Dayu and Zhao, Wenbo and Zhang, Hongbo and Chen, Zhang and Zeng, Wei and Wu, Wenbin},
copyright = {2024 Elsevier Ltd},
issn = {0306-4549},
journal = {Annals of nuclear energy},
language = {eng},
pages = {111067-},
publisher = {Elsevier Ltd},
title = {Research and application of the neutronics and thermal–hydraulic coupling based on the MORE framework},
volume = {212},
year = {2025},
}

@article{SchmidtRodney2015Aafc,
author = {Schmidt, Rodney and Belcourt, Kenneth and Hooper, Russell and Pawlowski, Roger and Clarno, Kevin and Simunovic, Srdjan and Slattery, Stuart and Turner, John and Palmtag, Scott},
copyright = {2014 Elsevier Ltd},
issn = {0306-4549},
journal = {Annals of nuclear energy},
language = {eng},
pages = {140-152},
publisher = {Elsevier Ltd},
title = {An approach for coupled-code multiphysics core simulations from a common input},
volume = {84},
year = {2015},
}

@article{FiorinaCarlo2015GanO,
author = {Fiorina, Carlo and Clifford, Ivor and Aufiero, Manuele and Mikityuk, Konstantin},
copyright = {2015 Elsevier B.V.},
issn = {0029-5493},
journal = {Nuclear engineering and design},
language = {eng},
pages = {24-37},
publisher = {Elsevier B.V},
title = {GeN-Foam: a novel OpenFOAM® based multi-physics solver for 2D/3D transient analysis of nuclear reactors},
volume = {294},
year = {2015},
}

@article{iapws97,
  author  = {Wagner, W. and Cooper, J. R. and Dittmann, A. and Kijima, J. and
             Kretzschmar, H.-J. and Kruse, A. and Mare{\v{s}}, R. and Oguchi, K.
             and Sato, H. and St{\"o}cker, I. and {\v{S}}ifner, O. and
             Takaishi, Y. and Tanishita, I. and Tr{\"u}benbach, J. and Willkommen, Th.},
  title   = {The {IAPWS} Industrial Formulation 1997 for the Thermodynamic
             Properties of Water and Steam},
  journal = {Journal of Engineering for Gas Turbines and Power},
  volume  = {122},
  number  = {1},
  pages   = {150--184},
  year    = {2000},
  doi     = {10.1115/1.483186}
}

@misc{BullerwellLanceEdward2020AMCS,
language = {eng},
publisher = {ProQuest Dissertations \& Theses},
title = {A Multiphysics Coupling Scheme between OpenMC and OpenFOAM for High-fidelity, Flexible Multi-physics Analysis of Nuclear Reactors},
year = {2020},
author = {Bullerwell, Lance Edward},
copyright = {Database copyright ProQuest LLC; ProQuest does not claim copyright in the individual underlying works.},
isbn = {9798607301453},
}

@techreport{godfrey2014vera,
  author      = {Godfrey, Andrew T.},
  title       = {{VERA} Core Physics Benchmark Progression Problem Specifications},
  institution = {Consortium for Advanced Simulation of Light Water Reactors (CASL), Oak Ridge National Laboratory},
  address     = {Oak Ridge, TN, USA},
  type        = {Technical Report},
  number      = {CASL-U-2012-0131-004},
  note        = {Revision 4},
  month       = aug,
  year        = {2014}
}

@article{kelly2017,
  author  = {Kelly, III, Daniel J. and Kelly, Ann E. and Aviles, Brian N. and
             Godfrey, Andrew T. and Salko, Robert K. and Collins, Benjamin S.},
  title   = {{MC21/CTF} and {VERA} multiphysics solutions to {VERA} core physics
             benchmark progression problems 6 and 7},
  journal = {Nuclear Engineering and Technology},
  volume  = {49},
  pages   = {1326--1338},
  year    = {2017},
  doi     = {10.1016/j.net.2017.07.016}
}

@techreport{fink1982uo2,
  author      = {Fink, J. K.},
  title       = {Tables of Thermodynamic and Transport Properties of {UO$_2$}},
  institution = {Argonne National Laboratory, Chemical Engineering Division},
  address     = {Argonne, IL, USA},
  number      = {ANL-CEN-RSD-82-3},
  month       = jun,
  year        = {1982}
}

@article{preCICEv2,
  author  = {Chourdakis, Gerasimos and Davis, Kyle and Rodenberg, Benjamin and Schulte, Miriam and Simonis, Fr{\'e}d{\'e}ric and Uekermann, Benjamin and Abrams, Georg and Bungartz, Hans-Joachim and Cheung Yau, Lucia and Desai, Ishaan and Eder, Konrad and Hertrich, Richard and Lindner, Florian and Rusch, Alexander and Sashko, Dmytro and Schneider, David and Totounferoush, Amin and Volland, Dominik and Vollmer, Peter and Koseomur, Oguz Ziya},
  title   = {{preCICE v2: A sustainable and user-friendly coupling library}},
  journal = {Open Research Europe},
  volume  = {2},
  pages   = {51},
  year    = {2022},
  doi     = {10.12688/openreseurope.14445.2}
}

@article{OpenFOAM,
  author  = {Weller, Henry G. and Tabor, Gavin and Jasak, Hrvoje and Fureby, Christer},
  title   = {{A tensorial approach to computational continuum mechanics using object-oriented techniques}},
  journal = {Computers in Physics},
  volume  = {12},
  number  = {6},
  pages   = {620--631},
  year    = {1998},
  doi     = {10.1063/1.168744}
}

@article{SchollSebastian2018Sosc,
journal = {Computers \& fluids},
language = {eng},
pages = {209-225},
publisher = {Elsevier Ltd},
title = {Stability of static conjugate heat transfer coupling approaches using Robin interface conditions},
volume = {172},
year = {2018},
author = {Scholl, Sebastian and Janssens, Bart and Verstraete, Tom},
address = {Amsterdam},
copyright = {2018 Elsevier Ltd},
issn = {0045-7930},
}

@inproceedings{HorelikN.2013Bfea,
language = {eng},
organization = {American Nuclear Society, 555 North Kensington Avenue, La Grange Park, IL 60526 (United States)},
publisher = {American Nuclear Society - ANS; La Grange Park (United States)},
title = {Benchmark for evaluation and validation of reactor simulations (BEAVRS)},
year = {2013},
author = {Horelik, N. and Herman, B. and Forget, B. and Smith, K.},
address = {United States},
}

@book{LuscherWalterG2010MPCC,
language = {eng},
publisher = {Pacific Northwest National Laboratory (U.S.)},
title = {Material Property Correlations: Comparisons between FRAPCON-3.4, FRAPTRAN 1.4, and MATPRO},
year = {2010},
author = {Luscher, Walter G and Geelhood, Kenneth J},
}

\end{document}